\documentclass{aa}
\usepackage{amsmath,graphicx,amssymb}
\usepackage[varg]{txfonts}
\usepackage{pst-eps}
\bibpunct{(}{)}{,}{a}{}{,}

\newcommand{\zav}[1]{\left(#1\right)}
\newcommand{\hzav}[1]{\left[#1\right]}
\newcommand{\szav}[1]{\left\{#1\right\}}
\newcommand{\prm}[1]{\langle#1\rangle}
\newlength\staretab
\newcommand{\Teff}{\mbox{$T_\mathrm{eff}$}}
\newcommand\de{\text{d}}
\newcommand\x[1]{\ensuremath{#1_\text{X}}}
\newcommand\lx{\ensuremath{\x L}}
\newcommand{\vel}{{\varv}}
\newcommand{\vnek}{{\varv_\infty}}
\newcommand\cpl{^+}
\newcommand\cmi{^-}
\newcommand\rhoc{{\rho\cpl}}
\newcommand\rhom{\rho\cmi}
\newcommand\cc{\ensuremath{C_\text{c}}}
\newcommand{\kms}{\ensuremath{\text{km}\,\text{s}^{-1}}}
\newcommand\ergs{\ensuremath{\text{erg}\,\text{s}^{-1}}}

\makeatletter
\def\sgn{\mathop{\operator@font sgn}\nolimits}
\makeatother

\begin{document}

\title{Wind inhibition by X-ray irradiation in HMXBs: the influence of clumping
and the final X-ray luminosity}

\author{J.~Krti\v{c}ka\inst{1} \and J. Kub\'at\inst{2} \and
I.~Krti\v{c}kov\'a\inst{1}}

\institute{\'Ustav teoretick\'e fyziky a astrofyziky, Masarykova univerzita,
           Kotl\'a\v rsk\' a 2, CZ-611\,37 Brno, Czech
           Republic
           \and
           Astronomick\'y \'ustav, Akademie v\v{e}d \v{C}esk\'e
           republiky, Fri\v{c}ova 298, CZ-251 65 Ond\v{r}ejov, Czech Republic}

\date{Received}

\abstract{In wind-powered X-ray binaries, the radiatively driven stellar wind
from the primary may be inhibited by the X-ray irradiation. This creates the
feedback that limits the X-ray luminosity of the compact secondary. Wind
inhibition might be weakened by the effect of small-scale wind inhomogeneities
(clumping) possibly affecting the limiting X-ray luminosity.}{We study the
influence of X-ray irradiation on the stellar wind for different radial
distributions of clumping.}{We calculate hot star wind models with external
irradiation and clumping using our global wind code. The models are calculated
for different parameters of the binary. We determine the parameters for which
the X-ray wind ionization is so strong that it leads to a decrease of the
radiative force. This causes a decrease of the wind velocity and even of the
mass-loss rate in the case of extreme X-ray irradiation.}{Clumping weakens the
effect of X-ray irradiation because it favours recombination and leads to an
increase of the wind mass-loss rate. The best match between the models and
observed properties of high-mass X-ray binaries (HMXBs) is derived with radially
variable clumping. We describe the influence of X-ray irradiation on the
terminal velocity and on the mass-loss rate in a parametric way. The X-ray
luminosities predicted within the Bondi-Hoyle-Lyttleton theory agree nicely with
observations when accounting for X-ray irradiation.}{The ionizing feedback
regulates the accretion onto the compact companion resulting in a relatively stable
X-ray source. The wind-powered accretion model can account for large
luminosities in HMXBs only when introducing the ionizing feedback. There are two
possible states following from the dependence of X-ray luminosity on the
wind terminal velocity and mass-loss rate. One state has low X-ray luminosity
and a nearly undisturbed wind, and the second state has high X-ray luminosity
and exhibits a strong influence of X-rays on the flow.}


\titlerunning{The influence of clumping on wind inhibition in HMXBs}

\authorrunning{J.~Krti\v{c}ka et al.}
\maketitle

\section{Introduction}

X-ray binaries contain a non-degenerate donor star, which deposits part of its
mass on the degenerate companion, typically a neutron star or a black hole. This
mechanism gives rise to objects which belong to the most luminous X-ray sources
in the universe. A class of high-mass X-ray binaries (HMXBs) is powered by
accretion of the radiatively driven wind blowing from the non-degenerate star
\citep{davos,laheupet}.

The wind in HMXBs has a very complex structure on various spatial scales. Accretion onto the compact companion proceeds in different forms depending on the
strength of the magnetic field of the compact star and on its rotation
\citep[see][for a review]{martpreh}. On large scales, the primary star wind is
strongly influenced by X-ray irradiation originating from the compact companion.
Since hot star winds are driven by light absorption in lines of heavier
elements, the strong X-ray irradiation affects the line force
\citep{sekerka,ff}. This results in a complex structure of the flow that has to
be studied using numerical simulations \citep{blondyn,felabon,hacek,elmel}. The
simulations predict the existence of a photoionization wake, which was detected
observationally \citep{huib}.

The complex structure of the flow causes severe problems in numerical
simulations. Self-consistent calculation of the radiative force requires
determination of ionization and excitation state of the flow, which is far from
equilibrium. Therefore, the radiative transfer equation has to be solved
together with equations describing ionization and excitation balance (kinetic
equilibrium equations). This resulting
NLTE
problem (that allows departures from
the   LTE) has not yet been solved in its full
complexity.

To make the problem tractable, the modelling either simplifies the radiative
driving enabling us to unveil the large-scale three-dimensional (3D) flow structure
\citep{blondyn}, or it solves the wind equations in spherical symmetry (1D) and
provides detailed calculation of the radiative force. The latter approach
\citep{velax1, dvojvit, sandvelax} is able to estimate binary parameters for
which the influence of X-rays becomes important. Wind models with detailed
radiative force also show the decrease of the wind velocity due to X-ray
irradiation, which was detected observationally \citep{vanlok,viteal}.

An overly strong X-ray source may even inhibit the wind before it reaches the
compact companion and quenches itself. As a result, there is a limit to the X-ray
luminosity, which depends on the properties of a given binary system. This can
be conveniently displayed in diagrams relating the X-ray luminosity and the
optical depth parameter \citep{dvojvit}. The positions of observed stars in such
diagrams appear in the region below the limit of wind inhibition by the X-ray
source in agreement with theory. Moreover, many stars appear close to the
border of wind inhibition indicating that their X-ray emission may be
self-regulated.

The impact of X-rays on the wind flow is also sensitive to the wind mass-loss
rate. Recombination is stronger for higher density (higher mass-loss rate),
reducing the effect of X-rays. Therefore, a possible reduction of wind mass-loss
rate estimates from observations \citep{dvojx31,bouhil,clres2,cohcar} may pose a
problem for wind models with X-ray irradiation. 
A similar issue also comes from wind modelling, because
the global (unified, i.e.
including the photosphere) hot star wind models predict mass-loss rates that are lower than the previous theoretical calculations
by a factor of between two and five
\citep{cmfkont}.
%

This problem can be alleviated by inclusion of clumping into wind models
\citep{osfek}. Clumping in hot star winds is connected with the appearance of
small-scale structures in the wind. This influences the ionization equilibrium
\citep{hamko,bourak,martclump,pulchuch} and the radiative transfer in the case
of optically thick clumps either in continuum \citep{lidarikala} or in lines
\citep{pof,lidaarchiv,chuchcar,sund,clres1,clres2}. Clumping is considered to be
one of the sources of X-ray variability in HMXBs
\citep[e.g.][]{prvni,osfek,manousci,bozof}. The origin of clumping is likely
the line-driven wind instability \citep{lusol,ornest}, which may be initiated
either in the photosphere by the turbulent motions \citep{felpulpal,cabra,jian}
or self-initiated in the wind \citep{sundsim}.

To understand the effect of clumping on the X-ray ionization, we included
clumping into our METUJE wind models \citep{cmfkont}, which  also include X-ray
irradiation. Inclusion of various physical effects into wind models requires
slight modifications of computational strategy. This is reflected in the structure of this paper. Section~\ref{global} describes the calculation
of global wind models, which neglect wind clumping and X-ray irradiation. The
flux from global models is subsequently used in the wind models with various
radial distributions of clumping (Sect.~\ref{chuch}). The inclusion of X-ray
irradiation, as well as wind models that include both clumping and X-ray irradiation, are
described in Sect.~\ref{irad}. The implications of our models for the X-ray
luminosity of HMXBs are discussed in Sect.~\ref{lxsec}.

\section{Global wind models}
\label{global}

The wind modelling is based on our METUJE code \citep{cmfkont}. The code
provides global (unified) photosphere-wind models. It solves the radiative
transfer equation, the kinetic (statistical) equilibrium equations, and the
hydrodynamic equations both in the photosphere and in the wind. The models are
calculated assuming that the flow is stationary (time-independent) and
spherically symmetric.

We solve the radiative transfer equation in the comoving-frame (CMF) following
the method developed by \citet{mikuh}. We include line and continuum transitions
relevant in atmospheres of hot stars in the radiative transfer equation. The
inner boundary condition for the radiative transfer equation is derived from the
diffusion approximation, and we assume neither an additional source of radiation
in the wind nor irradiation of the outer boundary.

The ionization and excitation states of considered elements \citep[see the
list in][]{nlteiii} are calculated from the kinetic equilibrium equations (also
known as NLTE equations). The equations account for the radiative and
collisional excitation, deexcitation, ionization, and recombination. A part of
the models of ions was adopted from the TLUSTY model stellar atmosphere input
data \citep{ostar2003,bstar2006}. Because the original TLUSTY ionic models are
tailored to stellar atmospheres, additional ionic models are needed for wind
modelling. We prepared these additional models using the data from the Opacity
and Iron Projects \citep{topt,zel0} and using the level data from NIST
\citep{nist}. For phosphorus the ionic model was prepared using data described
by \citet{pahole}. The bound-free radiative rates are consistently calculated
from the CMF mean intensity, while for the bound-bound rates we still use the
Sobolev approximation.

To derive the temperature, we use a differential form of the transfer equation
deep in the photosphere, while we use an integral form of this equation in the
upper layers of the photosphere \citep{kubii}. The electron thermal balance
method \citep{kpp} is applied in the wind. The hydrodynamical equations, that
is, the continuity equation, equation of motion, and the energy equation, are
solved iteratively. From this we obtain the wind density, velocity, and
temperature structure. The radiative force due to line and continuum transitions
is calculated in the CMF. The line data for the calculation of the line force
were taken from the VALD database (Piskunov et al. \citeyear{vald1}, Kupka et
al. \citeyear{vald2}) with some updates using the NIST data \citep{nist}.

\begin{table}
\caption{Adopted parameters of the model grid (stellar effective temperature
$\Teff$, radius $R_{*}$, mass $M_*$, and luminosity $L$).}
\centering
\label{ohvezpar}
\begin{tabular}{lc@{\hspace{3mm}}c@{\hspace{3mm}}r@{\hspace{3mm}}c@{\hspace{2mm}}c}
\hline
\hline
&Model &$\Teff$ & $R_{*}$ & $M_*$ & $\log(L/L_\odot)$ \\
&& $[\text{K}]$ & $[\text{R}_{\odot}]$ & $[\text{M}_{\odot}]$ \\
\hline Supergiants
& 300-1 & 30000 & 22.4 & 28.8 & 5.56  \\
& 375-1 & 37500 & 19.8 & 48.3 & 5.84  \\ Main-sequence
& 300-5 & 30000 &  6.6 & 12.9 & 4.50  \\
& 375-5 & 37500 &  9.4 & 26.8 & 5.19  \\
\hline
\end{tabular}
\end{table}

We calculated wind models for stellar parameters that roughly correspond to
typical parameters of O star primaries in HMXBs. We selected two supergiants and
two main sequence stars with $\Teff=30\,000 \,\text{K}$ and $\Teff=37\,500
\,\text{K}$. Stellar masses and radii of these stars given in
Table~\ref{ohvezpar} were derived using relations of \citet{okali} for main
sequence stars and supergiants. We assumed solar chemical composition
\citep{asp09} for our models.

\section{Wind models with optically thin clumping}
\label{chuch}

\subsection{Simplification of wind models}
\label{sanctasimplicitas}

The calculation of global wind models is relatively time consuming. Luckily, the
photospheric turbulence does not significantly modify the emergent flux.
Therefore, to make the following calculations more tractable, we do not account
for the photosphere in the wind models with clumping. Instead, we use the global
models to calculate the flux, which we subsequently apply as the lower boundary
flux in our wind models with clumping. A similar approach was employed in our
previous models that used the core-halo approximation \citep{velax1}. The flux
in the global models depends on radius, and we selected such a  flux, which leads to
a mass-loss rate close to that of the global models.

Because our CMF procedure of the calculation of the radiative force can be used
only for monotonic flows and the flow in the presence of the external
irradiation may become non-monotonic, we do not use a CMF calculation of the
radiative force directly in the following models. Instead, we calculate the
ratio of the CMF and Sobolev line force from a wind model without clumping and
use this ratio to correct the Sobolev line force in the models with clumping and
subsequently also in models with external X-ray irradiation. The Sobolev line
force is calculated from actual level populations \citep[][Eq.~25]{nltei}, which
may be influenced by clumping and X-ray irradiation (see Sect.~\ref{irad}).

\subsection{General assumptions}

The self-consistent way of including clumping into our wind models would
require employment of the time-dependent hydrodynamical simulations
\citep{ornest,felpulpal,owpu,runow,luk2d,felto,sundsim}. However, the
self-consistent solution of both NLTE equations and equations of hydrodynamics
is likely beyond the possibilities of contemporary computers. To make the task
more tractable, we include an 
approximate
effect of local wind inhomogeneities (clumps)
into our models.

The basic assumptions under which the clumps are included into our wind models
are the following:
\begin{enumerate}
\item The whole wind material is concentrated into the spatially
organized structures (clumps), the space between individual clumps is void.
The clumps are distributed randomly.
\item Each clump is homogeneous.
\item The presence of clumping explicitly affects only the wind density. The
wind velocity is assumed to be smooth and a monotonically
increasing function of radius.
\item We assume that the clumps
are optically thin both in continuum and in lines. This option is also referred
to as micro-clumping, that is, the opposite of macro-clumping which accounts for clumps that
may be optically thick \citep{lidarikala,lidaarchiv}.
\end{enumerate}
Under these assumptions the wind density $\rho$ is given by
\begin{equation}
\label{rhor}
\rho=\left\{
  \begin{array}{l}
    \rhoc\quad\text{in the clumps,}\\
    \rhom\quad\text{outside the clumps,}
  \end{array}\right.
\end{equation}
where $\rhom=0$ according to our assumptions. We note that $\rhoc$ is a function of
radius. Following \citet{abich}, we introduce the filling factor $f$ as a
probability that a given volume element is situated in the clump. The mean
quantities are calculated over the volume comprising a large number of clumps. In fact, it would be possible to introduce two types of averaging,
that is, the volume one,
which is defined
for a given instant of time, and the average over a
sufficiently long time interval. However, for simplicity, we assume that both
approaches are equivalent.

According to these assumptions, the mean wind density $\prm\rho$ 
\begin{equation}
\label{rhos}
\prm\rho=f\rhoc,
\end{equation}
gives the mass-loss rate that does not explicitly depend on $f$.
We note that $\prm\rho$ is also a function of radius. The mean value
of the density squared is
\begin{equation}
\label{drhos}
\prm{\rho^2}=f\rhoc^2.
\end{equation}
Following \citet{peichuch} for example, we introduce the clumping factor as
\begin{equation}
\label{chuchde}
\cc=\frac{\langle\rho^2\rangle}{\langle\rho\rangle^2}.
\end{equation}
Using Eqs.~\eqref{rhos} and \eqref{drhos} it can be shown that
for the case of void space between clumps the clumping factor is the
inverse of the filling factor,
\begin{equation}
\cc=\frac{1}{f}
,\end{equation}
\citep[see also][where the clumping factor is denoted as $D$]{hamko}.

The radially dependent mean density $\langle\rho\rangle$ is the same in the
smooth wind as in the clumped wind with the same mass-loss rate (and velocity
profile). Consequently, the explicit form of all equations that involve the
linear terms in density remains the same. The possibility of clumping is
included only in the terms in which the density squared features. In these term
it is necessary to formally substitute $\rho^2\rightarrow\prm{\rho^2}
\rightarrow\cc\prm{\rho}^2$ (see Eq.~\eqref{chuchde} and,
e.g.~\citealt{abich,igor,schm,hamko}). The details of the implementation of
clumping into our wind code are given in the Appendix.

\subsection{Calculated models}
\label{kapchuchmod}

We studied two different radial stratifications of clumping. We assumed either
constant clumping factor or we adopted the empirical
radial clumping stratification from
\citet{najradchuch}
\begin{equation}
\label{najc}
\cc(r)=C_1+(1-C_1)e^{-\frac{\vel(r)}{C_2}}+(C_4-C_1)
e^{\frac{\vel(r)-\vnek}{C_3}}.
\end{equation}
Here $\vel(r)$ is the wind radial velocity, $\vnek$ is the wind terminal
velocity, $C_1$ and $C_4$ are the clumping factors close to the star and at
large distances from the star, respectively, and $C_2$ and $C_3$ define the
radial variations of clumping factor; $C_2$ sets the onset of clumping (in
velocity space) and $C_3$ determines the typical velocity at which the clumping
factor gradually changes from $C_1$ to $C_4$. Because Eq.~\eqref{najc} may give
$\cc<1$, we used a lower limit to \cc\ equal to 1. In comparison with monotonic
increase of the clumping factor with $r$, as adopted in some models
\citep[e.g.][]{bouhil}, Eq.~\eqref{najc} accounts for the decrease of clumping
in outer regions of dense winds \citep{pulchuch}. Motivated by typical values
derived in \citet{najradchuch,dvojx31}, we assume $C_1=10$, $C_2=200\,\kms$,
$C_3=500\,\kms$, and $C_4=1$.

\begin{figure}[t]
\centering
\resizebox{\hsize}{!}{\includegraphics{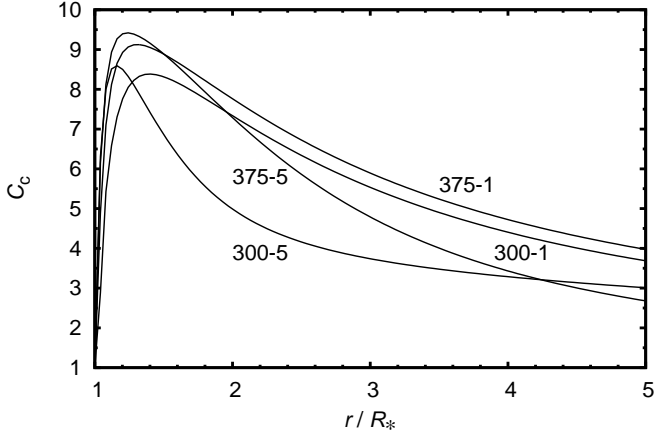}}
\caption{Radial variation of the clumping factor according to Eq.~\eqref{najc}
for individual wind models from Table~\ref{ohvezpar}.}
\label{najcobr}
\end{figure}

The dependence of clumping on actual wind velocity in Eq.~\eqref{najc} is not
practical for wind simulations. Therefore, within our calculations, we derive
\cc\ using the fit to the wind velocity predicted for smooth wind with $\cc=1$
(see Eq.~\eqref{vrfit}). Given the empirical nature of Eq. \eqref{najc} with
many uncertainties, this change is of minor importance. The radial variations
of \cc\ for individual wind models are given in Fig.~\ref{najcobr}.

\begin{table}[t]
\caption{Derived wind mass-loss rate $\dot M$,
wind terminal velocity $v_\infty$, and velocity law
fit parameters $v_1$, $v_2$, $v_3$, and $\gamma$ for models with
clumping. We specify either \cc\ for the
models with constant clumping factor or
$C_1$ for radially dependent clumping factor Eq.~\eqref{najc}.}
\label{chuchpar}
\centering
\begin{tabular}{c@{\hspace{2mm}}l@{\hspace{2mm}}c@{\hspace{2mm}}c@{\hspace{2mm}}c@{\hspace{2mm}}r@{\hspace{2mm}}r@{\hspace{2mm}}r}
\hline
Model & Clumps & $\dot M$ & $v_\infty$ &
$v_1$ & \multicolumn{1}{c}{$v_2$} & 
\multicolumn{1}{c}{$v_3$} & \multicolumn{1}{c}{$\gamma$}\\
&&$[\text{M}_{\odot}\,\text{yr}^{-1}$]& \multicolumn{4}{c}{[\kms]}\\
\hline
300-1 & $\cc=1$ & $4.5\times10^{-7}$ & 1510 & 2130 & 0 & $-510$ & 750\\
      & $\cc=3$ & $7.7\times10^{-7}$ & 1130 & 1930 & $-740$ & 0 & 1500\\
      & $C_1=10$& $6.6\times10^{-7}$ & 1480 & 2700 & $-1170$ & 0 & 1500\\
375-1 & $\cc=1$ & $1.3\times10^{-6}$ & 2020 & 3360 & $-1310$ & 0 & 1720\\
      & $\cc=3$ & $1.9\times10^{-6}$ & 1990 & 3030 & $-990$ & 0 & 9900\\
      & $C_1=10$& $1.7\times10^{-6}$ & 2470 & 4370 & $-1850$ & 0 & 1300\\
300-5 & $\cc=1$ & $1.3\times10^{-8}$ & 1630 & 5490 & $-6800$ & 3020 & 5600\\
      & $\cc=3$ & $3.7\times10^{-8}$ & 1430 & 2780 & $-1350$ & 0 & 14000\\
      & $C_1=10$& $1.5\times10^{-8}$ & 1340 & 3580 & $-4200$ & 2100 & 12000\\
375-5 & $\cc=1$ & $1.1\times10^{-7}$ & 2360 & 4620 & $-2280$ & 0 & 8700\\
      & $\cc=3$ & $2.1\times10^{-7}$ & 2470 & 4310 & $-1770$ & 0 & 9900\\
      & $C_1=10$& $2.0\times10^{-7}$ & 2750 & 5310 & $-2480$ & 0 & 8500\\
\hline
\end{tabular}
\end{table}

The basic parameters of the wind with clumping are given in
Table~\ref{chuchpar}. Clumping favours recombination, which leads to an increase of
the radiative force and therefore of the mass-loss rate, because ions with lower
charge drive the wind more efficiently \citep[e.g.][]{sanya}. For supergiants
we predict on average $\dot M\sim\cc^{0.4}$ for constant clumping factor, while
\citet{muij} predict $\dot M\sim\cc^{0.2-0.4}$  , in reasonable agreement. The
increase of the mass-loss rate is weaker in the models with radially variable
clumping factor after Eq.~\eqref{najc}, because the mass-loss rate is determined
by conditions close to the star in our models where clumping is weak (see
Fig.~\ref{najcobr}).

We fitted the radial velocity dependence $v(r)$ of these models using an
analytic formula (we denote the analytic fit by $\tilde v(r)$)
\begin{multline}
\label{vrfit}
\tilde \varv (r)=
\hzav{\varv_1\zav{1-\frac{R_*}{r}}+\varv_2\zav{1-\frac{R_*}{r}}^2+
\varv_3\zav{1-\frac{R_*}{r}}^3}\\\times
\szav{1-\exp\hzav{-\gamma\zav{1-\frac{r}{R_*}}^2}},
\end{multline}
where $\varv_1$, $\varv_2$, $\varv_3$, and $\gamma$ are parameters of the fit
given in Table~\ref{chuchpar} \citep[see also][]{betyna}. The radial increase of
clumping after Eq.~\eqref{najc} leads to an increase of the line force in the
outer wind, which may result in higher wind terminal velocity than in the models
with constant clumping factor. Therefore, radially increasing the clumping factor
can account for high observed terminal velocities, which are underestimated in
global models without clumping \citep{cmfkont}.

We also used the models to derive mean mass-absorption coefficient averaged
over radii $1.5\,R_*-5\,R_*$ and approximated its wavelength dependence as
\begin{equation}
\label{kapafit}
\log\zav{\frac{\tilde \kappa_\nu^\text{X}}{1\,\text{cm}^2\,\text{g}^{-1}}}=
  \left\{\begin{array}{l}
    \min(a_1\log\lambda+b_1,\log a_0),
    \quad \lambda<\lambda_1,\\
    a_2\log\lambda+b_2,\quad \lambda>\lambda_1,\\
  \end{array}\right.\\
\end{equation}
where $\lambda_1=20.18$, and $\lambda$ is the value of the wavelength in units
of \AA. The parameters of the fit $a_0$, $a_1$, $b_1$, $a_2$, and $b_2$ are
given in Table~\ref{fitko}. 
The resulting values of $\tilde \kappa_\nu^\text{X}$ correspond typically
within 10--20\% to those given in the literature \citep[e.g.][]{xlida}.
It follows that the coefficients of the fit do not
significantly depend on clumping in most cases; consequently, clumping does not
significantly affect the opacity in the X-ray domain \citep[see][]{lojza}. This
assumption is used to determine mass-loss rates from X-ray line profiles
\citep{cohcar}.

\begin{table}[t]
\caption{Derived fit coefficients of the mass-absorption coefficient
Eq.~\ref{kapafit} for individual models.}
\label{fitko}
\centering
\begin{tabular}{clccccc}
\hline
Model & Clumps & $a_0$ & $a_1$ & $b_1$ & $a_2$ & $b_2$\\
\hline
300-1 & $\cc=1$ & 205 & 2.618 & $-0.911$ & 2.634 & $-1.431$\\
      & $\cc=3$ & 205 & 2.616 & $-0.901$ & 2.617 & $-1.392$\\
      & $C_1=10$& 205 & 2.616 & $-0.900$ & 2.616 & $-1.390$\\
375-1 & $\cc=1$ & 180 & 2.512 & $-0.827$ & 2.590 & $-1.503$\\
      & $\cc=3$ & 190 & 2.569 & $-0.878$ & 2.637 & $-1.501$\\
      & $C_1=10$& 190 & 2.584 & $-0.892$ & 2.646 & $-1.499$\\
300-5 & $\cc=1$ & 155 & 2.361 & $-0.702$ & 1.886 & $-0.786$\\
      & $\cc=3$ & 190 & 2.612 & $-0.914$ & 2.658 & $-1.504$\\
      & $C_1=10$& 175 & 2.485 & $-0.799$ & 2.579 & $-1.503$\\
375-5 & $\cc=1$ & 165 & 2.361 & $-0.703$ & 1.778 & $-0.601$\\
      & $\cc=3$ & 185 & 2.508 & $-0.826$ & 2.580 & $-1.500$\\
      & $C_1=10$& 180 & 2.521 & $-0.836$ & 2.598 & $-1.505$\\
\hline
\end{tabular}
\end{table}

\section{Wind models with X-ray irradiation}
\label{irad}

\begin{figure}[t]
\centering
\resizebox{\hsize}{!}{\includegraphics{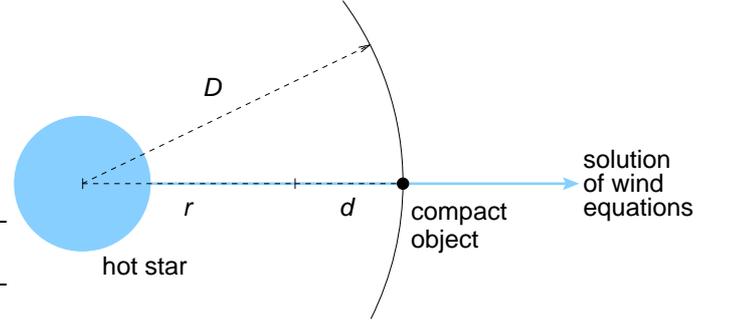}}
\caption{Geometry of the model of a hot star wind
irradiated by X-rays from a compact companion.
Here $D$ is binary separation, $r$ is radius of studied point in the wind, and
$d$ its distance from the compact companion.}
\label{geom}
\end{figure}

\begin{figure*}[t]
\centering
\resizebox{0.49\hsize}{!}{\includegraphics{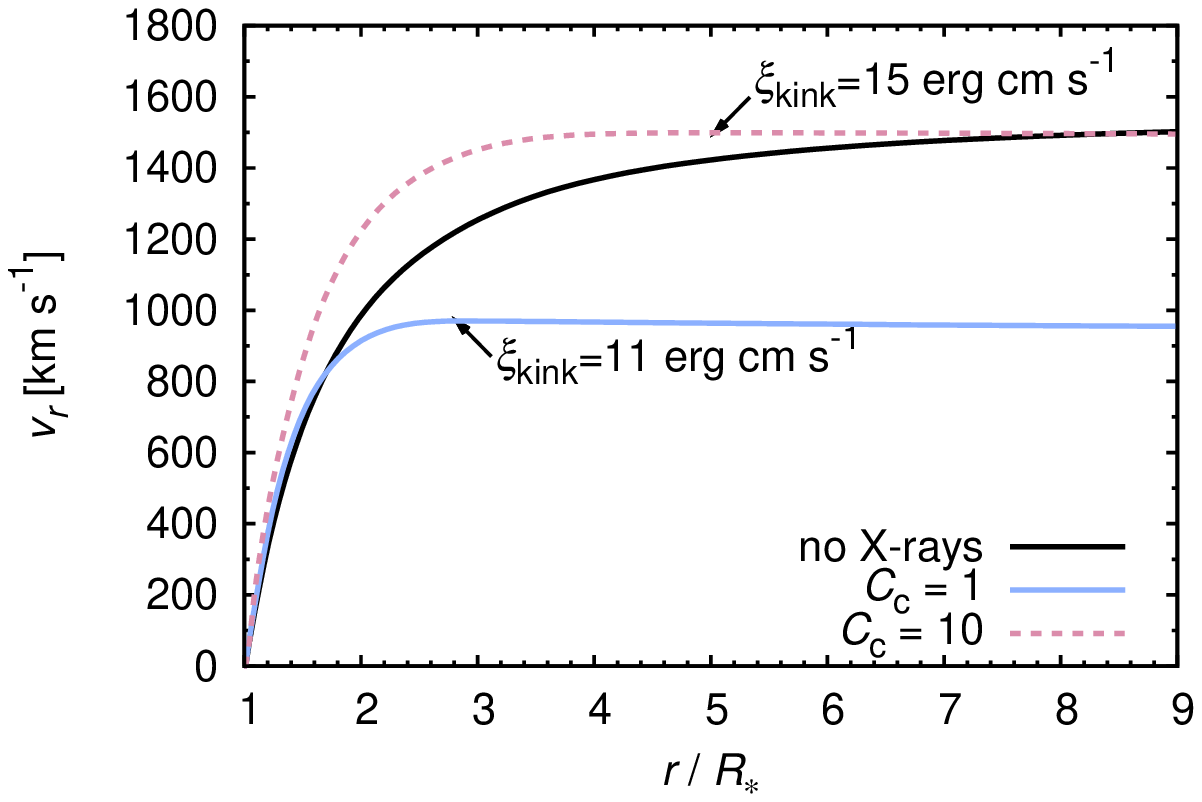}}
\resizebox{0.49\hsize}{!}{\includegraphics{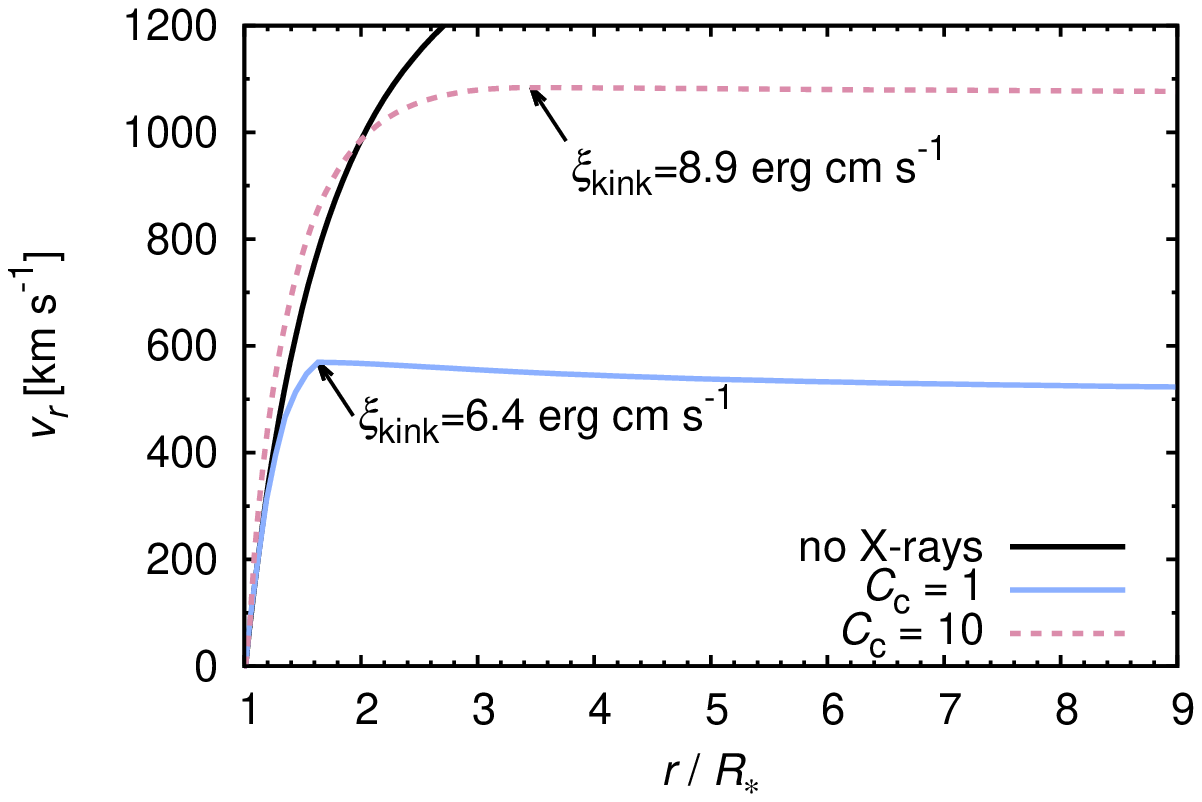}}
\caption{Plot of the radial velocity for selected wind models 300-1.
The black line denotes the model without either
X-ray irradiation or clumping, while blue and
red lines denote models with the same X-ray irradiation (with the same \lx\
and $D$) and different clumping: without clumping (blue
solid) and with $\cc=10$ (red dashed).
%
Here the clumping is allowed to affect
the ionization structure only directly and not via the mean density
(mass-loss rate).
Overplotted is the value of the ionization parameter Eq.~\eqref{xic} at the
location of the velocity kink.
{\em Left panel}: Model with $\lx=10^{36}\,\ergs$ and $D=200\,R_\odot$.
{\em Right panel}: Model with $\lx=10^{38}\,\ergs$ and $D=700\,R_\odot$.}
\label{vrobrjenxv}
\end{figure*}

\begin{figure*}[t]
\centering
\resizebox{0.49\hsize}{!}{\includegraphics{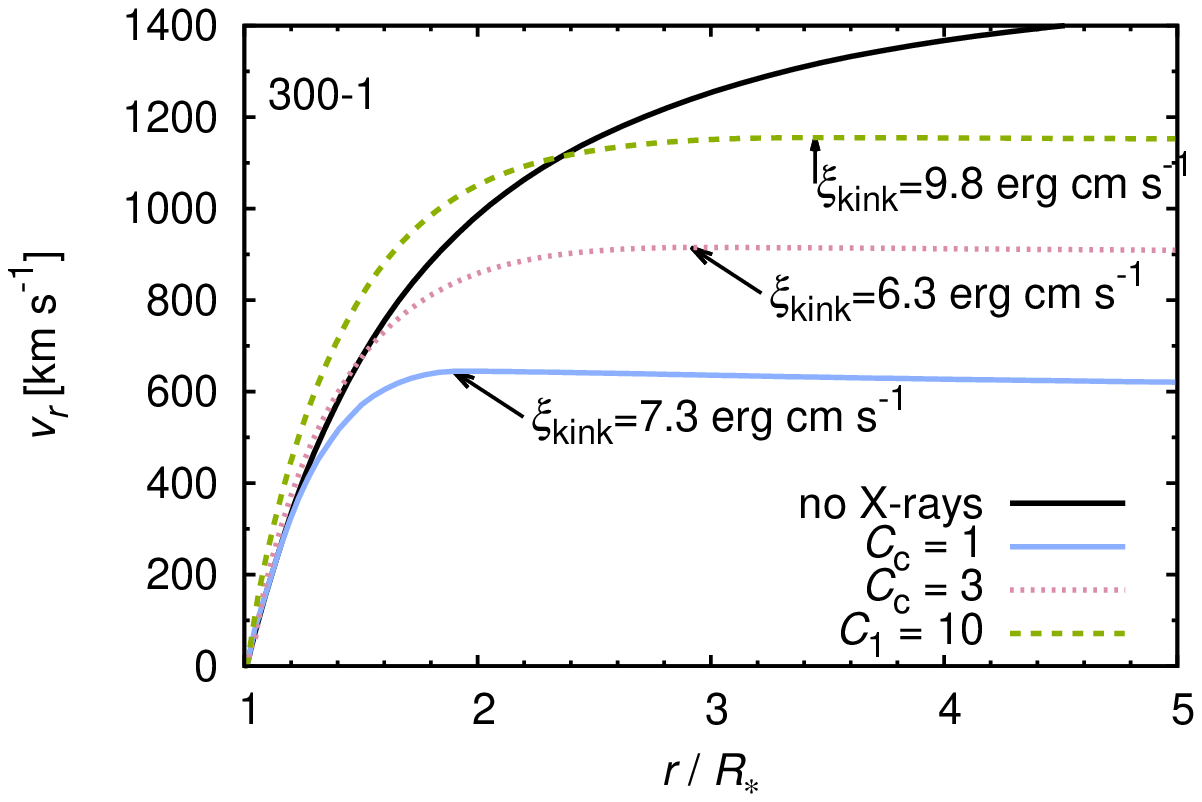}}
\resizebox{0.49\hsize}{!}{\includegraphics{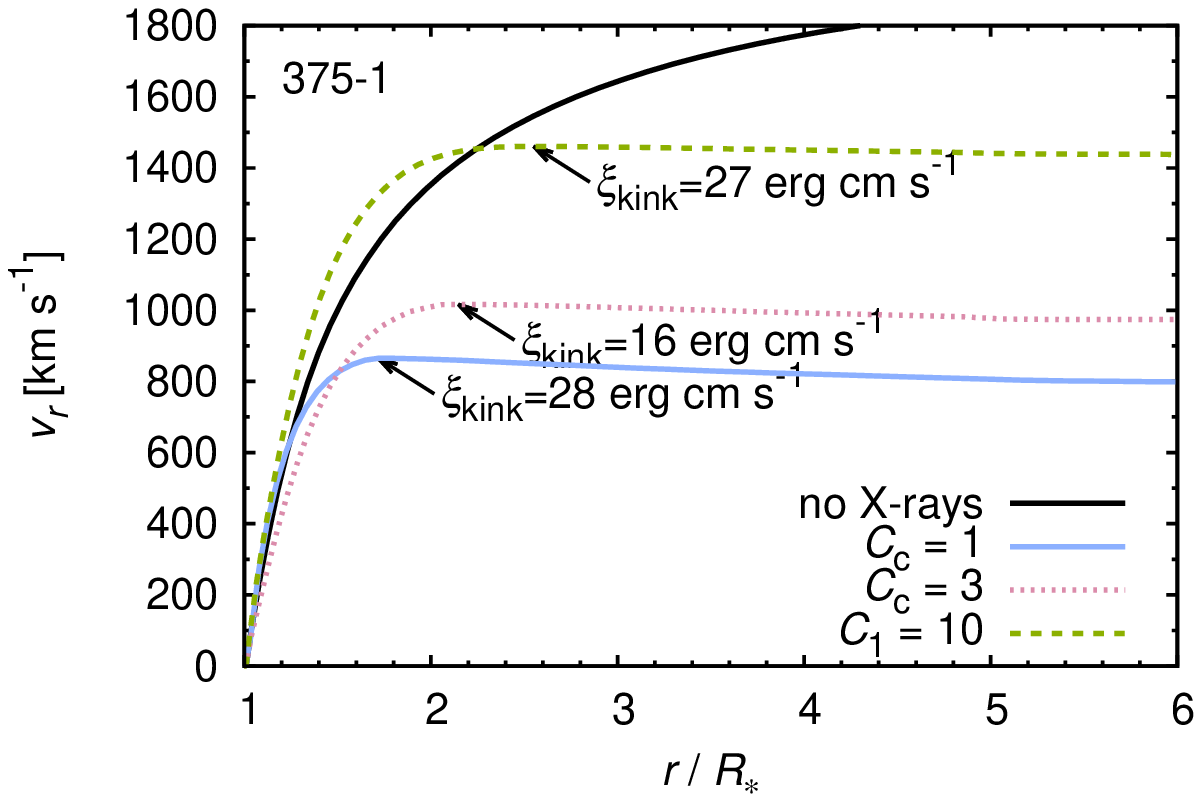}}
\resizebox{0.49\hsize}{!}{\includegraphics{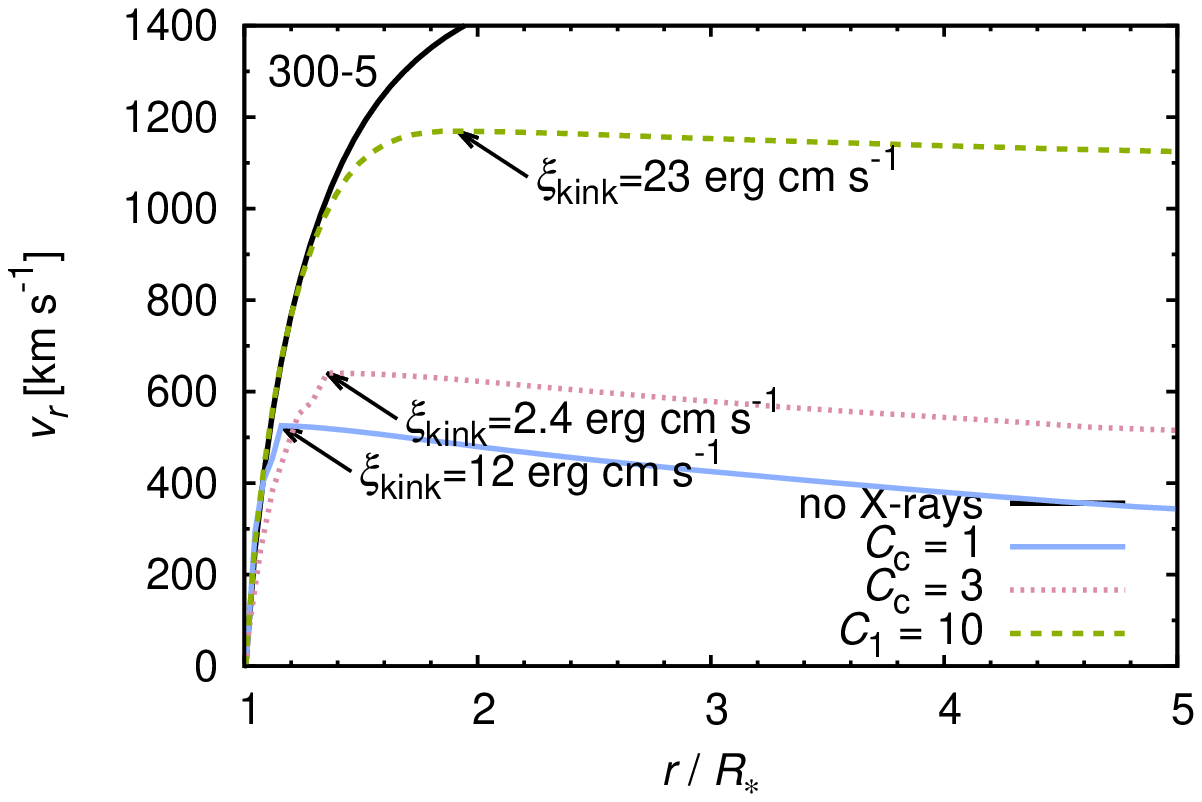}}
\resizebox{0.49\hsize}{!}{\includegraphics{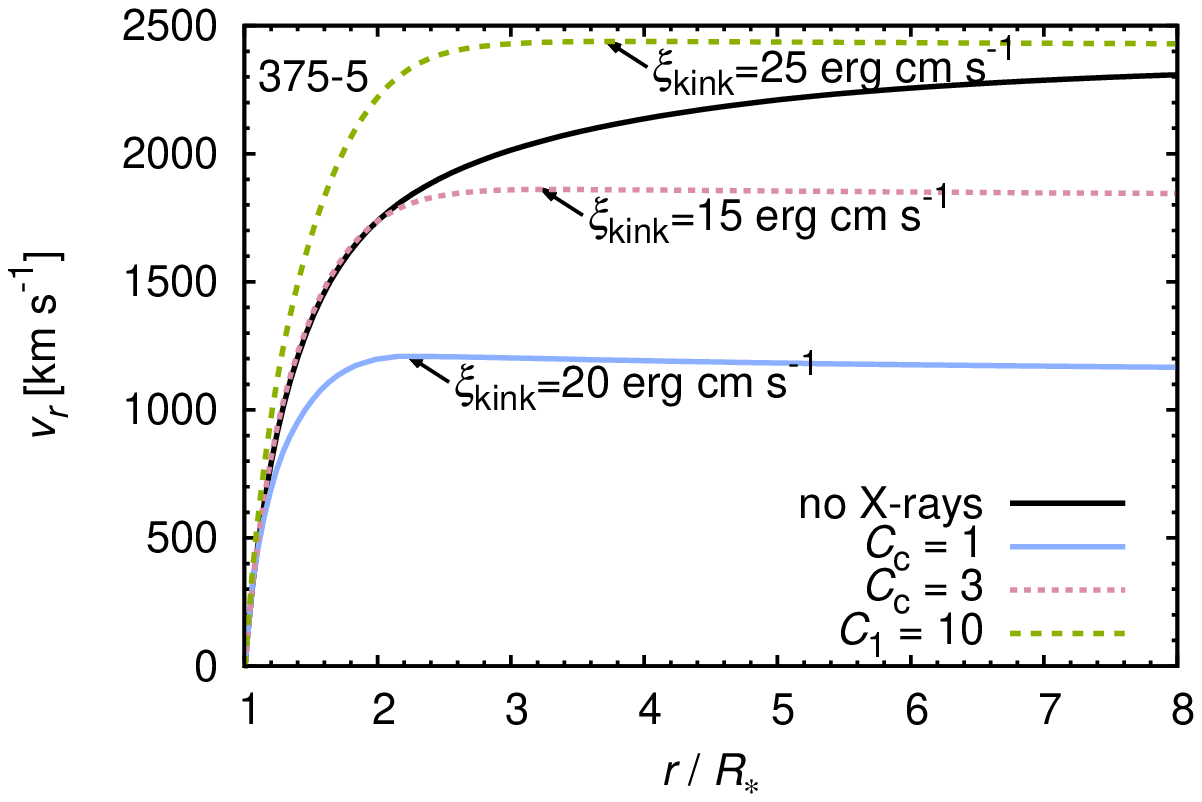}}
\caption{Plot of the radial velocity for selected wind models. The black line
denotes the model without either
X-ray irradiation or clumping, while blue, red, and green
lines denote models with the same X-ray irradiation (with the same \lx\ and $D$)
but with different clumping: without clumping (blue,
solid), with $\cc=3$ (red dotted), and
with radially dependent clumping factor after Eq.~\eqref{najc} with $C_1=10$
(green dashed). Arrows denote the location of the kink in the velocity profile with
a corresponding value of the ionization parameter  from Eq.~\eqref{xic}.
{\em Upper left}: Model 300-1 with $\lx=10^{37}\,\ergs$ and $D=300\,R_\odot$.
{\em Upper right}: Model 375-1 with $\lx=10^{37}\,\ergs$ and $D=100\,R_\odot$.
{\em Lower left}: Model 300-5 with $\lx=10^{35}\,\ergs$ and $D=30\,R_\odot$.
{\em Lower right}: Model 375-5 with $\lx=10^{36}\,\ergs$ and $D=100\,R_\odot$.}
\label{vrobr}
\end{figure*}

The inclusion of X-ray irradiation into our wind models closely follows
\cite{velax1}. In the present study, we aim at the most significant influence of
the X-ray irradiation, therefore we solve the wind equation only along a radial
ray in the direction of the X-ray source (see Fig.~\ref{geom}). Similarly as for
other models with clumping (described in Sect.~\ref{chuch}), we use the
photospheric flux and the Sobolev line force corrected for CMF radiative
transfer to solve wind equations (see Sect.~\ref{sanctasimplicitas}). Our
treatment of the Sobolev line force neglects non-local radiative coupling between
absorption zones, which occurs in non-monotonic flows \citep{rybashumrem,pof}.

The influence of the compact secondary component is only taken into account by
inclusion of external X-ray irradiation, which originates in the wind accretion
on the compact component. The X-ray irradiation is introduced as an additional
term in the mean intensity $J_\nu$
\begin{equation}
\label{xneutron}
J_\nu^\text{X}=\frac{L_\nu^\text{X}}{16\pi^2d^2}\text{e}^{-\tau_\nu(r)},
\end{equation}
where $L_\nu^\text{X}$ is monochromatic X-ray irradiation luminosity,
which after the integration gives the total X-ray luminosity, $\x L=\int
L_\nu^\text{X}\,\de\nu$. The frequency distribution of emergent X-rays
$L_\nu^\text{X}$ is for simplicity approximated by the power law
$L_\nu^\text{X}\sim\nu^{-1}$ from 0.5 to 20~keV \citep [see] []{viteal}.  The
distance of a given point in the wind from the compact companion is $d=|D-r|$ 
from Fig.~\ref{geom}, $D$ is binary separation, and $\tau_\nu(r)$ is the
optical depth between the given point in the wind and the compact companion,
\begin{equation}
\label{zamlhou}
\tau_\nu(r)=\left|\int_r^D\kappa_\nu(r')\rho(r')\,\de r'\right|,
\end{equation}
where $\kappa_\nu$ is the X-ray mass-absorption coefficient.

The absorption coefficient and the density in Eq.~\eqref{zamlhou} can be derived
directly from the actual model. However, we used a simplified approach to avoid
possible problems with the convergence of the model. Therefore, for the
calculation of the external X-ray irradiation in Eq.~\eqref{xneutron} we use the
optical depth from Eq.~\eqref{zamlhou} calculated with density, velocity, and
mass-absorption coefficient in the form of
\begin{equation}
\begin{split}
\rho(r)&=\frac{\dot M}{4\pi r^2 v(r)},\\
\varv(r)&=\min(\tilde \varv(r),\varv_\text{kink}),\\
\kappa_\nu(r)&=\tilde \kappa_\nu^\text{X},
\end{split}
\end{equation}
where $\varv_\text{kink}$ is the velocity of the kink that may appear in the
wind model (otherwise $\varv_\text{kink}=\infty$), $\tilde \varv(r)$ is the fit
\eqref{vrfit} of the wind velocity derived from the models without X-ray
irradiation (if the fit $\tilde \varv(r) > \varv_\text{kink}$ then the value of
$\varv_\text{kink}$ is taken), and $\tilde \kappa_\nu^\text{X}$ is the radially
averaged mass-absorption coefficient  given by Eq.~\eqref{kapafit}.

\subsection{Influence of clumping on the X-ray wind inhibition}

\newrgbcolor{sarlatova}{0.79 0.37 0.37}
\newrgbcolor{plosar}{0.86 0.55 0.67}
\newrgbcolor{nejakazelena}{0.20 0.64 0.65}

\begin{figure*}[t]
\centering
\resizebox{0.448\hsize}{!}{\includegraphics{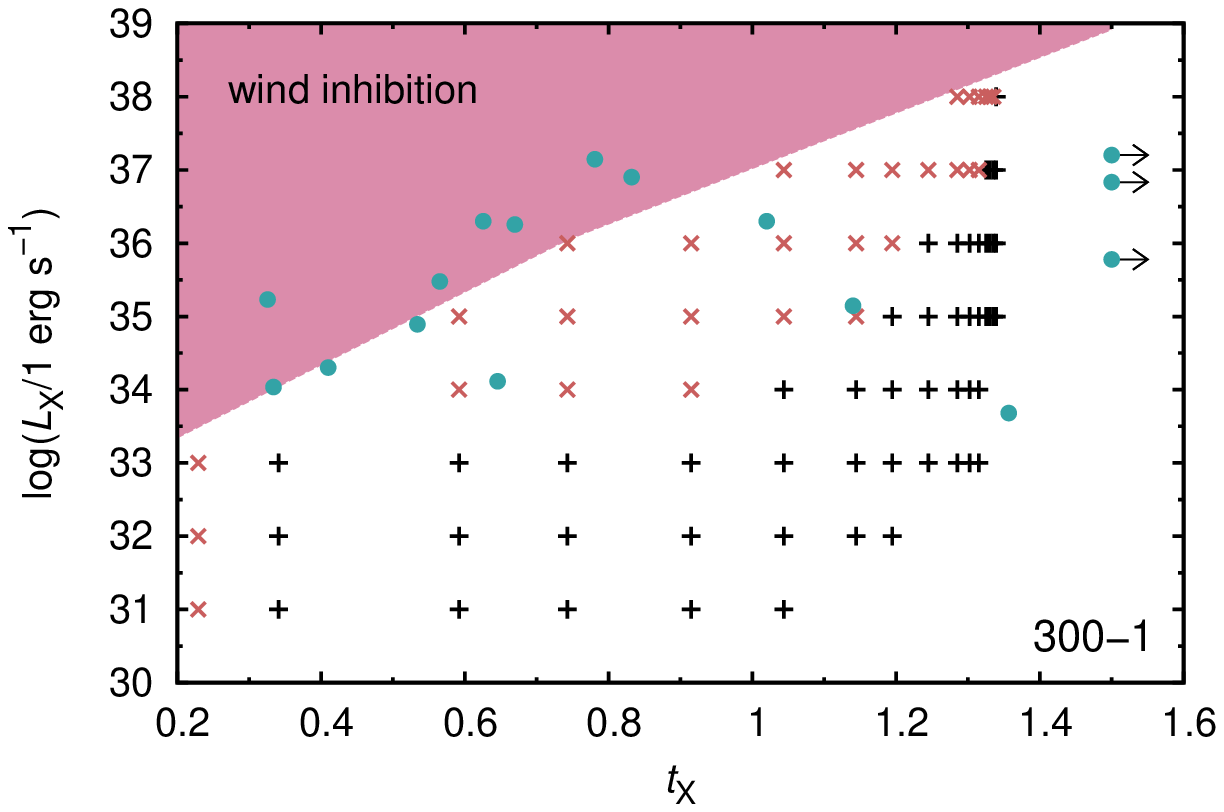}}
\resizebox{0.448\hsize}{!}{\includegraphics{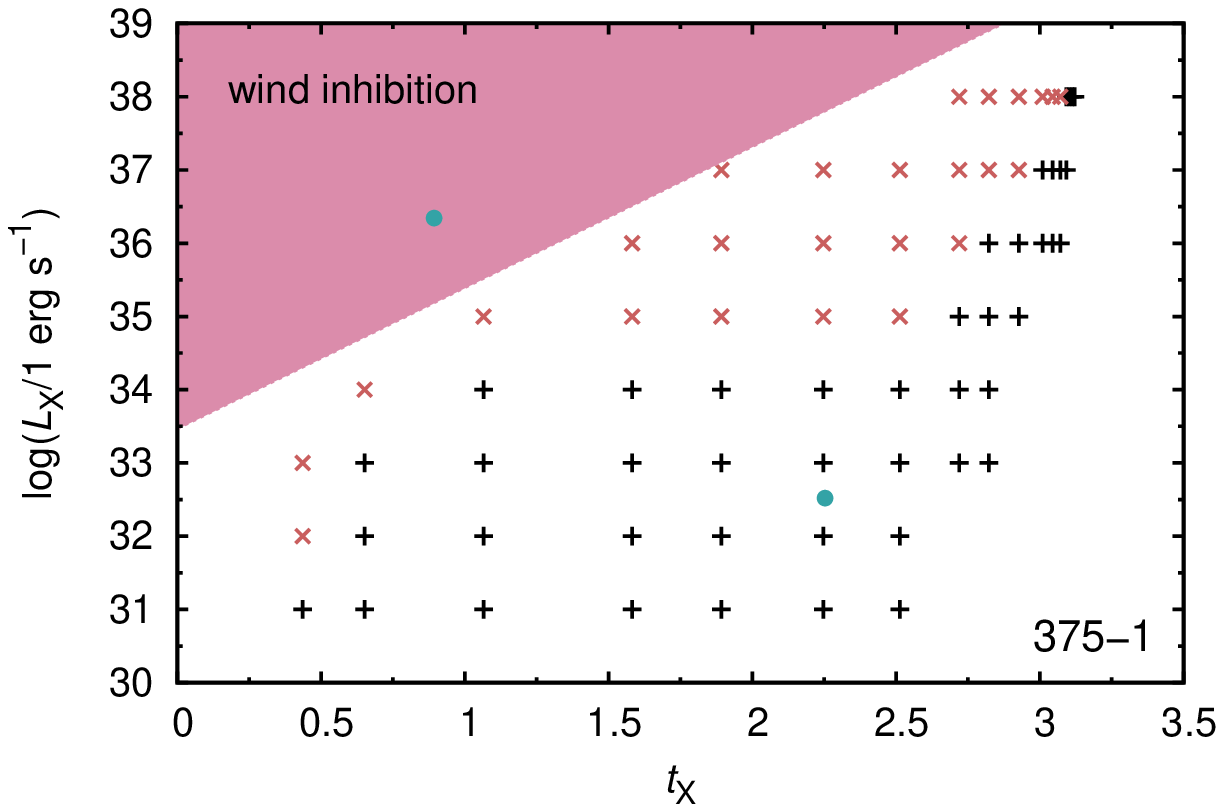}}
\resizebox{0.448\hsize}{!}{\includegraphics{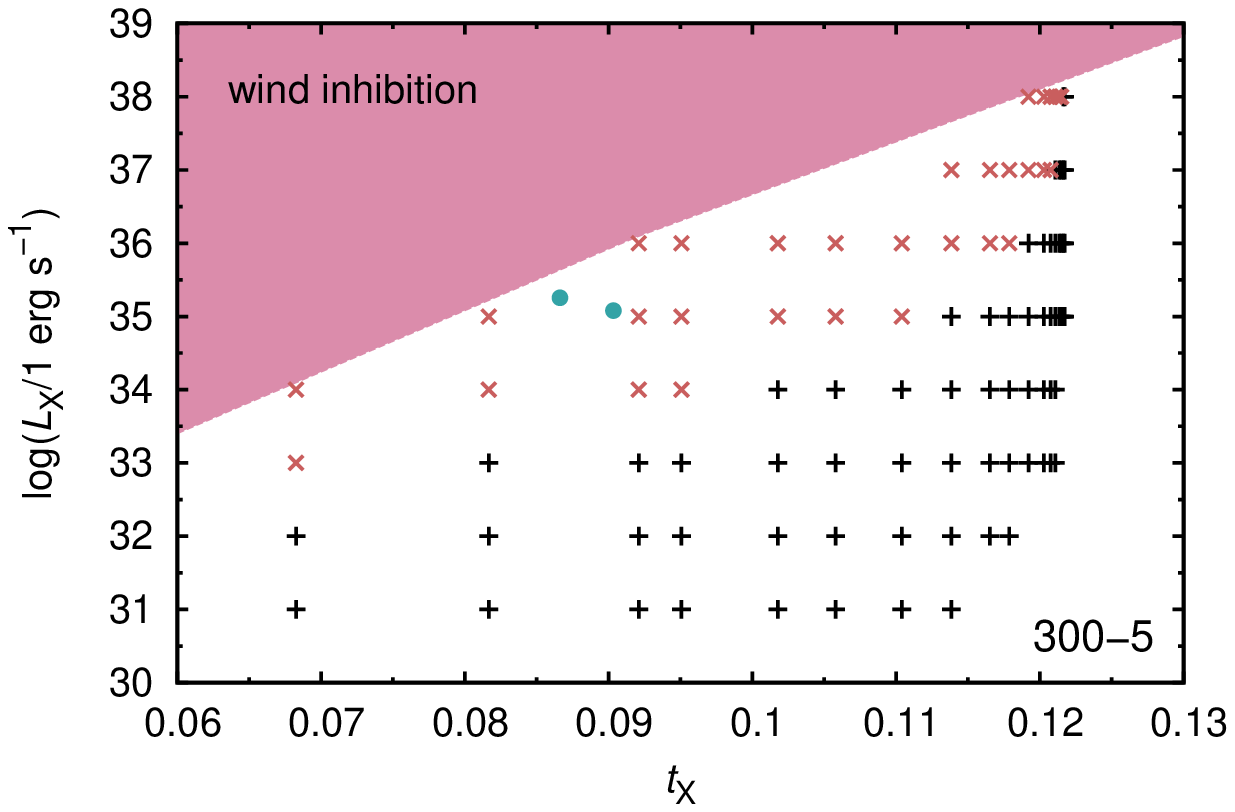}}
\resizebox{0.448\hsize}{!}{\includegraphics{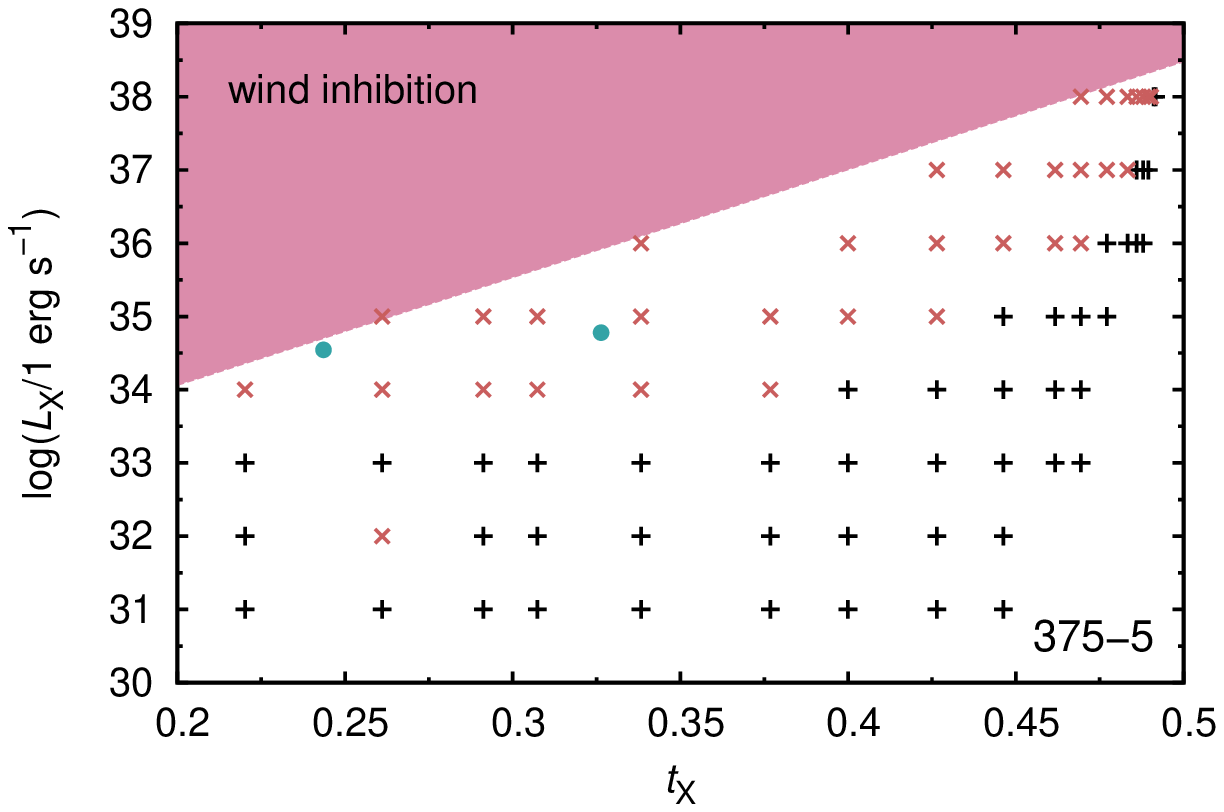}}
\caption{Diagrams of X-ray luminosity vs.~the optical depth parameter
Eq.~\eqref{tx} for models without clumping and for individual stars from
Table~\ref{ohvezpar}. Each point describes models with different \lx\ and $D$.
Individual symbols discriminate between different effects of X-ray ionization on
the wind: black plus symbols ($\boldsymbol+$) denote models with negligible
influence of X-ray irradiation, while red crosses
({\sarlatova$\boldsymbol\times$}) denote models where X-ray irradiation leads to
the decrease of the wind terminal velocity. The shaded area
({\plosar\protect\raisebox{-1.5pt}{\protect\rule{8pt}{8pt}}}) marks the regions of the
$\lx-\x t$ parameters where the wind inhibition appears. Overplotted are the
positions of non-degenerate components of HMXBs from Table~\ref{neutron} (filled
circles, {\nejakazelena\Large\raise -1pt\hbox{\textbullet}}).}
\label{lxtx}
\end{figure*}

\begin{figure*}[p]
\centering
\resizebox{0.448\hsize}{!}{\includegraphics{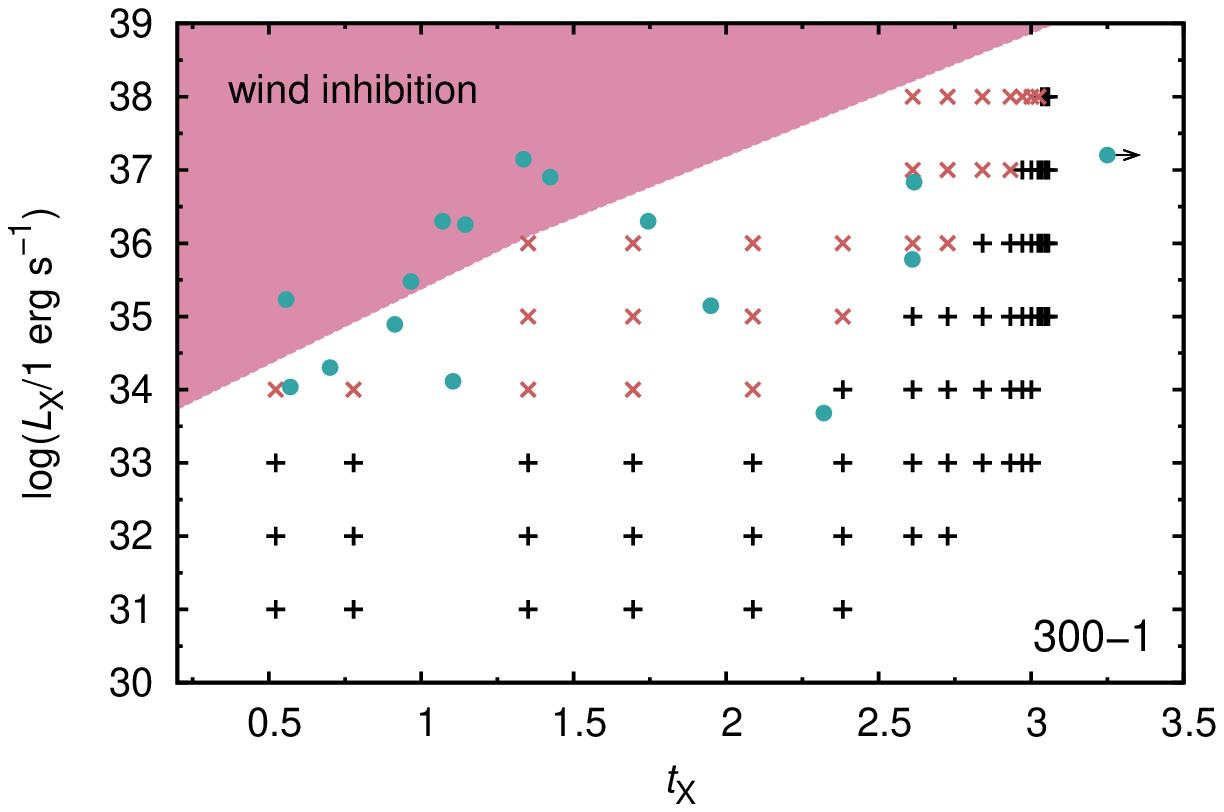}}
\resizebox{0.448\hsize}{!}{\includegraphics{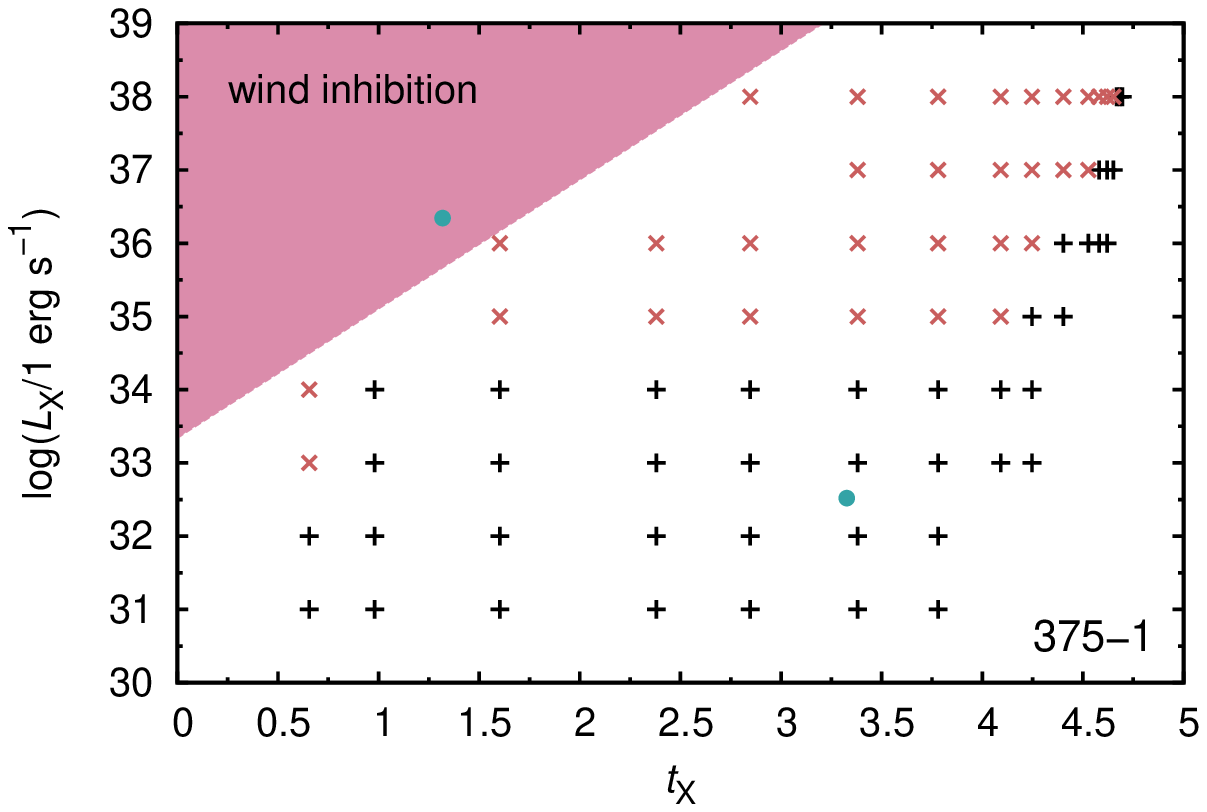}}
\resizebox{0.448\hsize}{!}{\includegraphics{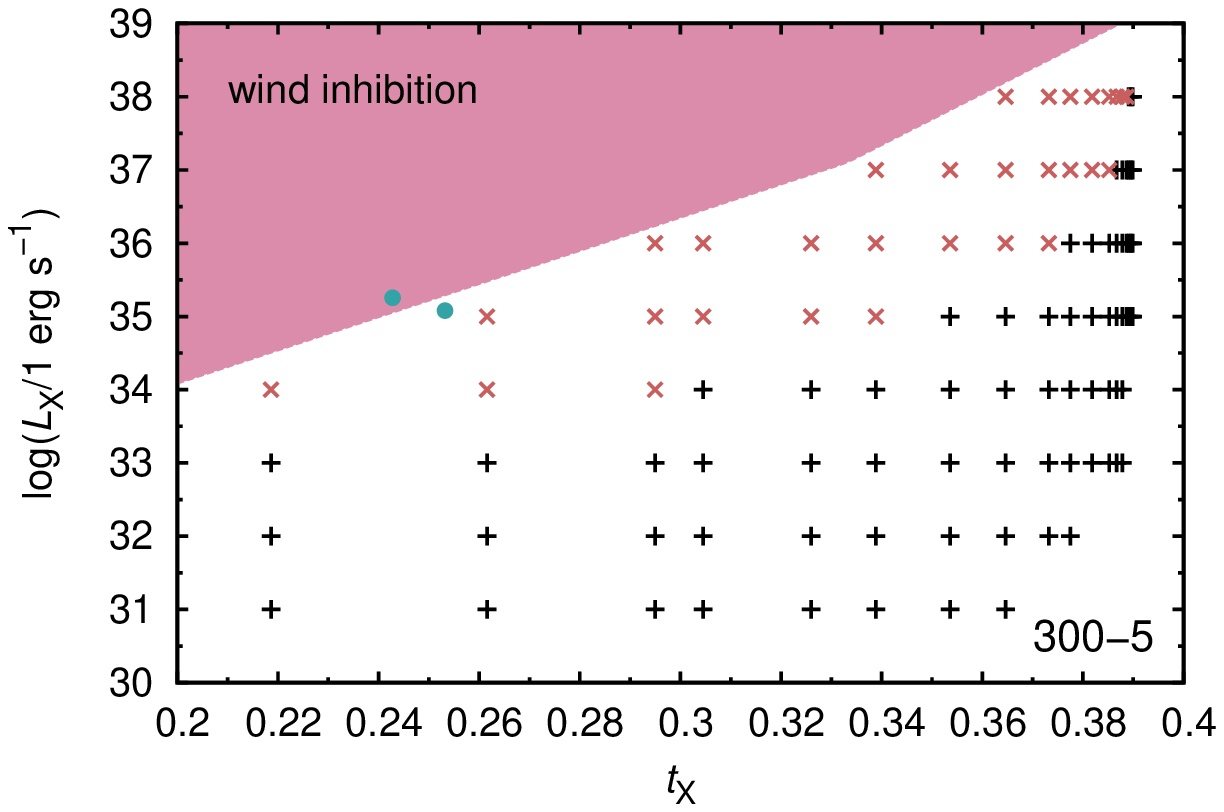}}
\resizebox{0.448\hsize}{!}{\includegraphics{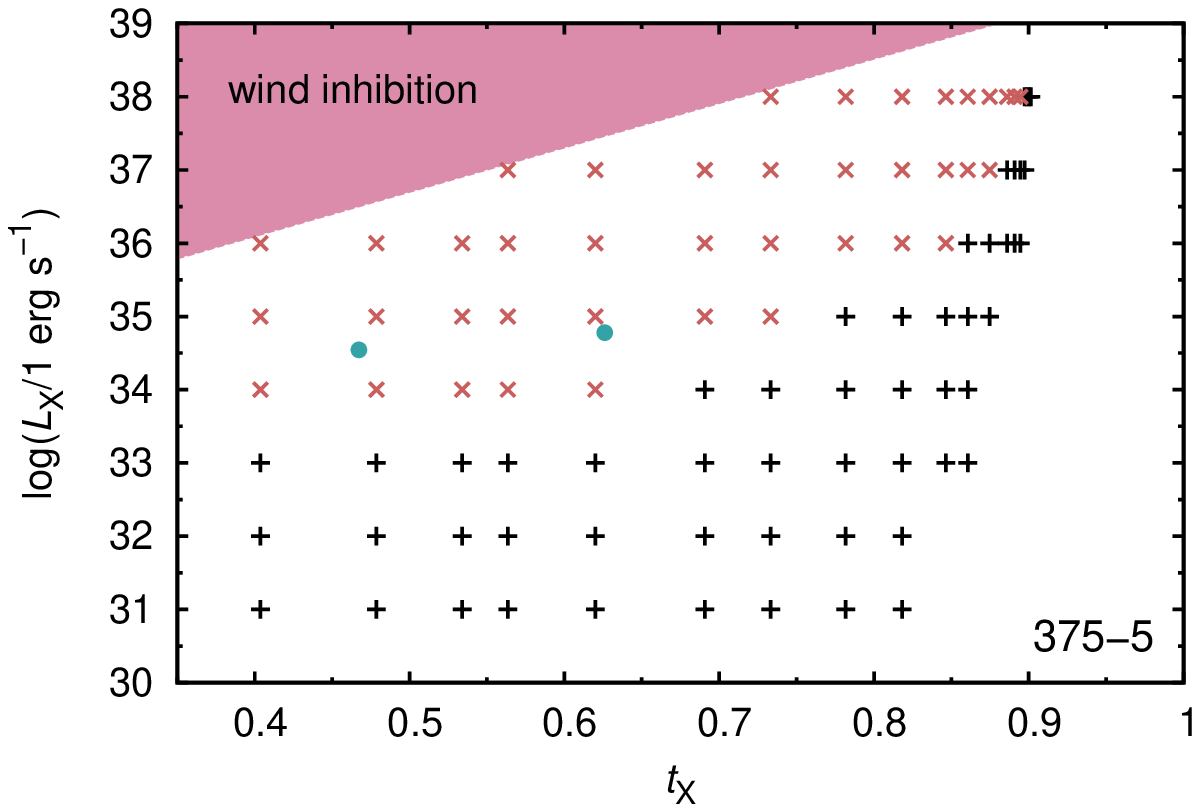}}
\caption{As in Fig.~\ref{lxtx}, but using a constant clumping factor
$\cc=3$}
\label{lxtxch3}
\end{figure*}

\begin{figure*}[p]
\centering
\resizebox{0.448\hsize}{!}{\includegraphics{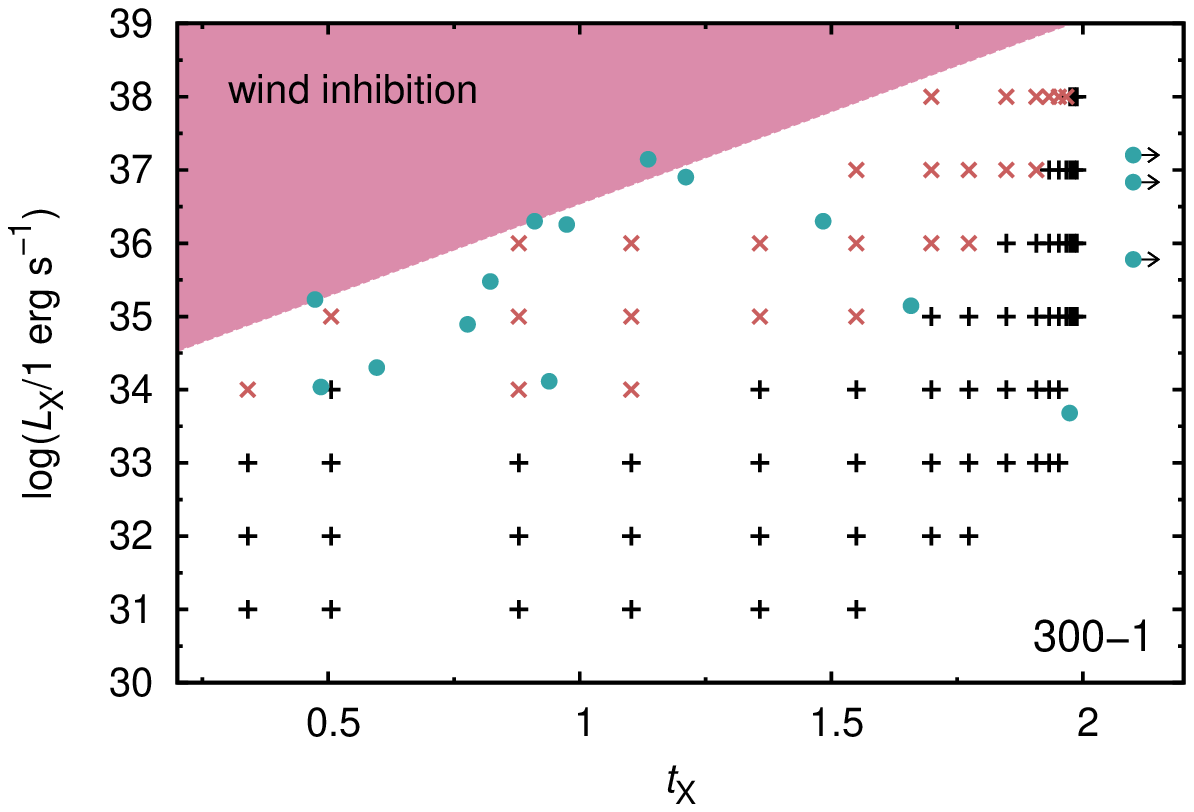}}
\resizebox{0.448\hsize}{!}{\includegraphics{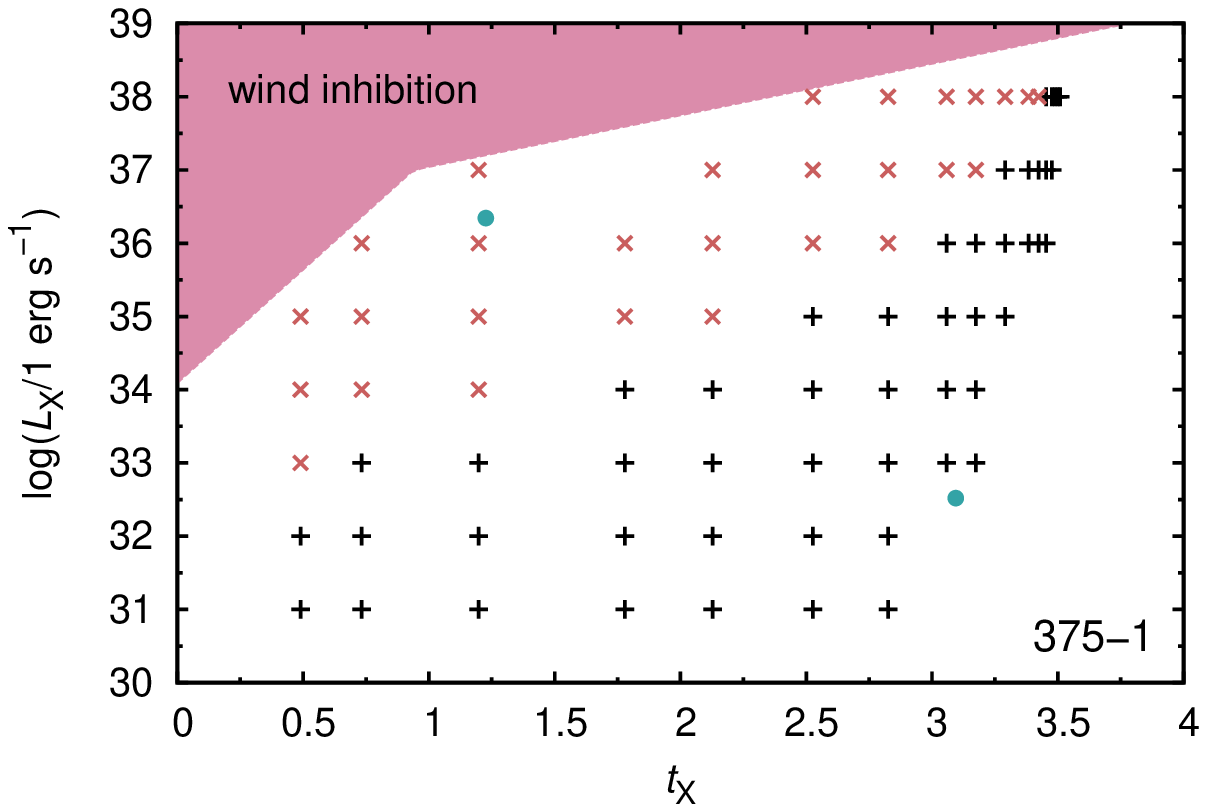}}
\resizebox{0.448\hsize}{!}{\includegraphics{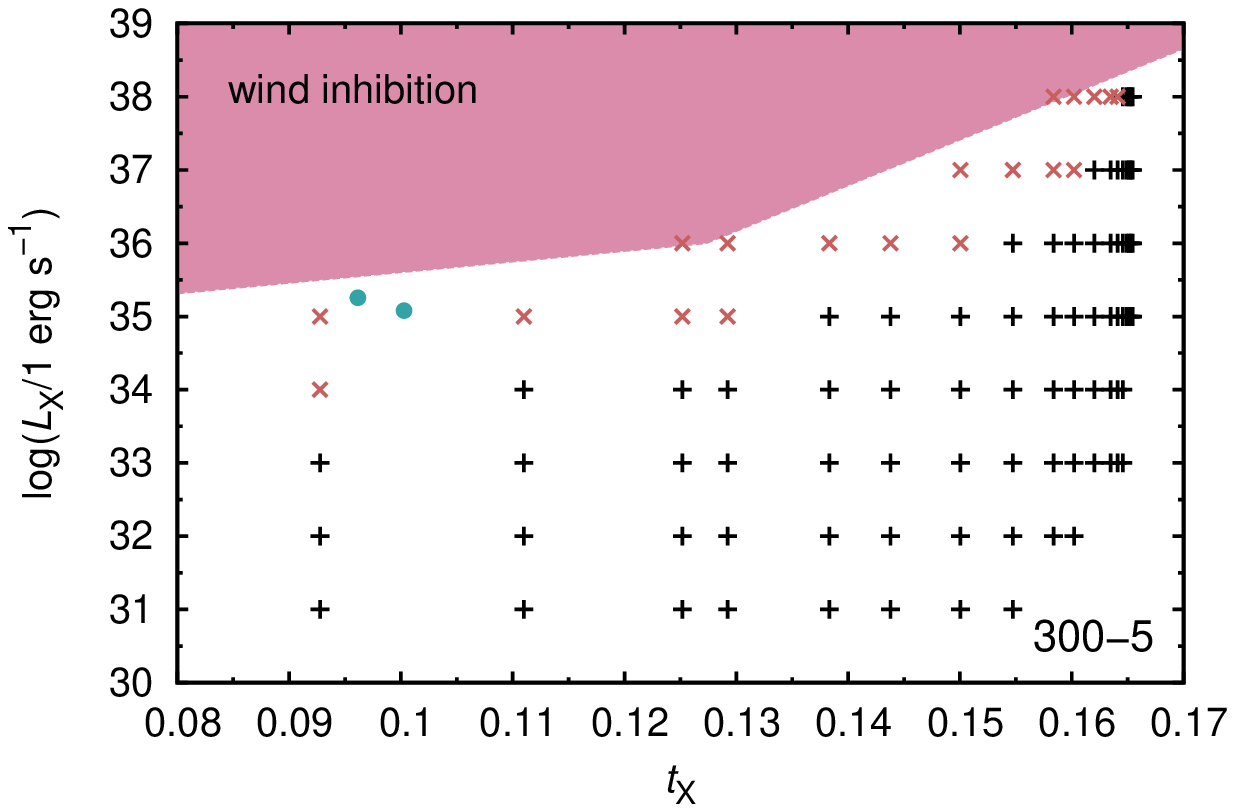}}
\resizebox{0.448\hsize}{!}{\includegraphics{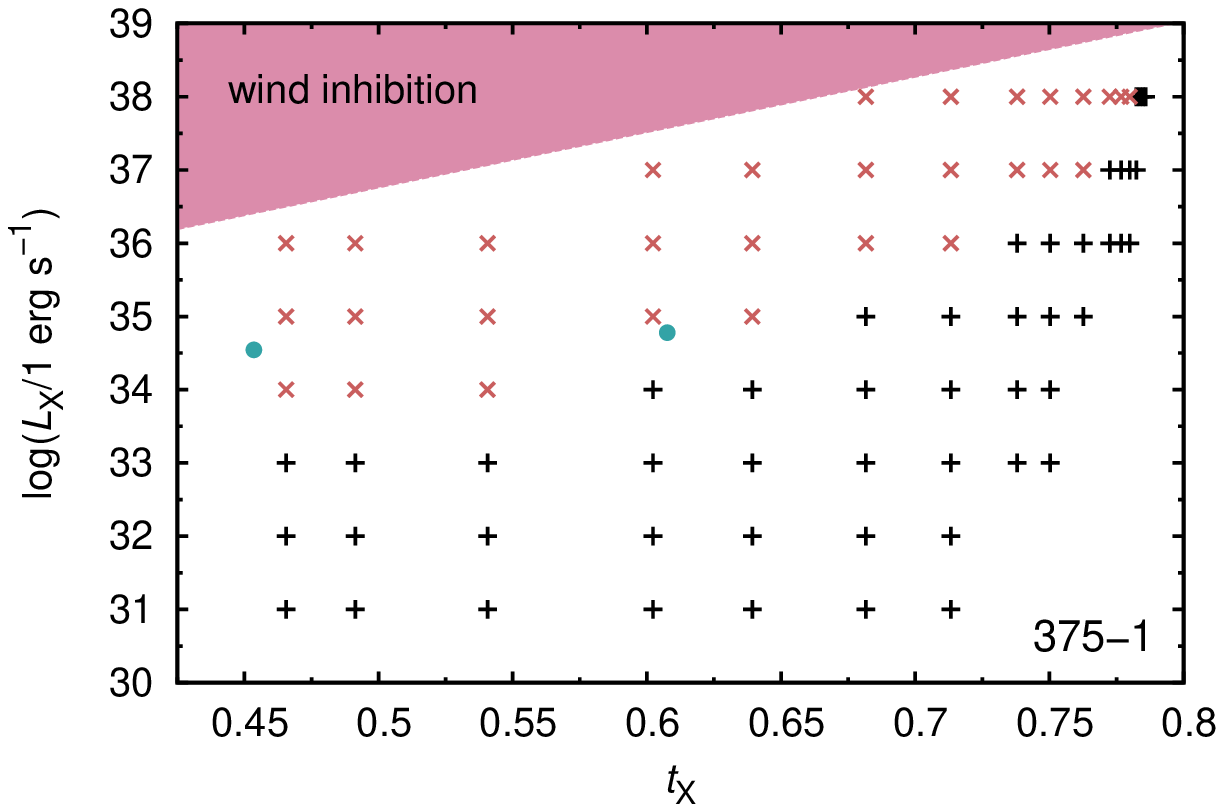}}
\caption{As in Fig.~\ref{lxtx}, but using the variable clumping factor from
Eq.~\eqref{najc} with $C_1=10$}
\label{lxtxch10naj}
\end{figure*}

The influence of X-rays on the ionization state of matter with atomic number
density $n$ is traditionally described using the ionization parameter
$\xi=\lx/(n d^2)$ introduced by \citet[][see also \citealt{sekerka}]{tenci}.
\citet{dvojvit} accounted for the X-ray absorption in the intervening
wind via the $\xi\sim e^{-\tau_\nu}$ dependence. There is an
additional effect in the presence of clumping, because a higher density of material inside the clumps enhances recombination without any impact
on photoionization. Therefore, one may
expect X-ray ionization to have  a weaker effect in clumped media, as shown by
\citet{osfek}. We introduce the ionization parameter as
\begin{equation}
\label{xic}
\xi(r)=\frac{1}{n d^2C_\text{c}}
\int L_\nu^\text{X}\text{e}^{-\tau_\nu(r)}\,\de\nu.
\end{equation}
From this we can infer two different ways by which optically thin clumps weaken the
effect of wind X-ray photoionization. The direct mechanism \citep{osfek} gives
$\xi\sim1/C_\text{c}$ due to the overdensity inside clumps which favours
recombination, while the indirect one is connected with variations of the wind
mass-loss rate and its influence on $n$ and $\tau_\nu$. Because the wind
mass-loss rate increases with increasing $C_\text{c}$ for optically thin clumps
\citep[see also Sect.~\ref{kapchuchmod}]{muij}, all these effects lead to a weaker
influence of X-rays on the clumped flow. Moreover, in winds which have high
mass-loss rates and which are optically thick in the X-ray domain
($\tau_\nu>1$), the indirect effect is expected to dominate due to the
exponential dependence on the optical depth in Eq.~\eqref{xic}.

To test the applicability of the ionization parameter~\eqref{xic} in the
case of clumping, we calculated wind models with constant clumping factor
$\cc=10$ where only the direct effect of clumping affects $\xi$. This can be
achieved only in the models that (without X-ray irradiation) give the same
density and velocity structure. To do so, we fixed the mass-loss rate and the
terminal velocity of the model without X-ray irradiation to values derived
without clumping (and without X-ray irradiation) by a multiplication of the
radiative force. The multiplicative factor was different below and above the
critical point, because we fixed both the mass-loss rate and the terminal
velocity. Thanks to this, we isolated the direct influence of clumping in
Eq.~\eqref{xic} and eliminated the indirect influence via $n$ and $\tau_\nu$
(see Fig.~\ref{vrobrjenxv}). For a low amount of X-ray irradiation or for a
distant X-ray source (low $\xi$), only wind velocity is affected by X-ray
irradiation \citep{dvojvit,sandvelax}. As a result of X-ray irradiation, higher
ionization states become more abundant. For weak X-ray irradiation, the lower
ionization states are not significantly affected; consequently, weak X-ray
irradiation causes a slight increase of the radiative force and of the wind
terminal velocity. However, a stronger X-ray irradiation (with larger $\xi$)
depopulates ions with low charge that mostly drive the wind, and the radiative
force and the terminal velocity decrease. This is typically accompanied by the
appearance of a kink in the velocity profile where the velocity derivative
changes sign. This is demonstrated in Fig.~\ref{vrobrjenxv}, which shows the
radial velocity in the models where clumping is allowed to affect the ionization
equilibrium only directly. The ionization parameter $\xi_\text{kink}$ at the
location of the kink is roughly the same in the models with and without
clumping, which demonstrates that Eq.~\eqref{xic} is able to reliably
characterize the influence of clumping on X-ray ionization.

From Fig.~\ref{vrobr} it follows that the ionization parameter \eqref{xic}
reasonably describes the effect of X-ray irradiation even in more realistic
models where the clumping affects both the ionization and wind structure. In
such models, both the direct and indirect effects of clumping in Eq.~\eqref{xic}
come into play. The plots in Fig.~\ref{vrobr} show the kink in the velocity
profile, where the wind is no longer accelerated due to X-ray ionization. For
a given star, Fig.~\ref{vrobr} shows that the kink appears for roughly the same
critical ionization parameter $\xi_\text {kink} \approx 5-25
\,\text{erg}\,\text{cm}\,\text{s}^{-1}$ in agreement with
\citet{dvojvit}\footnote{There we erroneously scaled down the value of the
ionization parameter (evaluated at the critical point) by a factor of
$(4\pi)^2$.}.

Figures~\ref{vrobrjenxv} and \ref{vrobr} also show that clumping weakens the
effect of X-ray irradiation. With stronger clumping, the kink in the velocity
profile appears at larger distances from the star and at lower wind densities
(higher velocities). This has important implications for the regions of wind
inhibition as they appear in diagrams of X-ray luminosity versus~the optical depth
parameter.

\subsection{Diagrams of X-ray luminosity versus~the optical depth parameter}

The diagrams that display X-ray luminosity versus~the optical depth parameter were
proven to be very effective in estimating the impact of X-rays on the wind
\citep{dvojvit}. The optical depth parameter,
\begin{equation}
\label{tx}
\x t=\frac{\dot M}{\varv_\infty}\zav{\frac{1}{R_*}-\frac{1}{D}}
\zav{\frac{10^3\,\text{km}\,\text{s}^{-1}\,1\,R_\odot}
{10^{-8}\,{M}_\odot\,\text{yr}^{-1}}}
,\end{equation}
is proportional to the radial optical depth between the stellar surface and the
X-ray source. Therefore, it enables us to separate domains according to the  type
of influence that X-rays have on the wind flow. We use the optical depth parameter
defined by Eq.~\eqref{tx} instead of the optical depth, because \x t\ does not
depend on energy. The value of $\x t$ in Eq.~\eqref{tx} is calculated with
mass-loss rate and terminal velocity unaffected by X-rays.

Diagrams of X-ray luminosity versus~the optical depth parameter are given in
Figs.~\ref{lxtx} -- \ref{lxtxch10naj} for different clumping factors (see
Table~\ref{chuchpar}). The wind is not affected by X-ray irradiation for low
\lx\ or for large binary separation. Such models appear in right bottom corner
of the $\lx-\x t$ diagrams as black plus symbols. With increasing \lx\ or with
decreasing binary separation, the influence of X-rays becomes stronger and leads
to the appearance of the kink in the velocity profile and to a decrease of the
terminal velocity marked by red crosses. For a large X-ray luminosity or for a
very close X-ray source, the kink approaches the wind critical point\footnote{The
critical point is defined as a point where the speed of the radiative-acoustic
Abbott waves is equal to the wind velocity \citep{abbvln,fero}.} leading to the
wind inhibition \citep{velax1}. Such models appear in the upper left corner of the
$\lx-\x t$ diagrams.

Figure~\ref{lxtx} displays the case without clumping ($\cc=1$). Adding the
constant clumping factor ($\cc>1$) in Fig.~\ref{lxtxch3} has two effects. The
mass-loss rate becomes higher, shifting all models in the diagram towards larger
$\x t$ , and the boundary marking the wind inhibition retreats towards larger \lx\
with respect to models without clumping. The latter effect once again
demonstrates the weakening of the effect of X-ray irradiation with increasing
clumping. Similar effects also appear in Fig.~\ref{lxtxch10naj} for the models
with the depth-variable clumping factor given by Eq.~\eqref{najc}. However, the shift
towards larger $\x t$ is shorter than in models with $\cc=3$ as a result of
the weaker increase of the mass-loss rate.

\begin{table*}[t]
\caption{Parameters of HMXBs with a neutron star or black hole companion.}
\label{neutron}
\centering
\begin{tabular}{l@{\hspace{2.5mm}}c@{\hspace{2.5mm}}c@{\hspace{2.5mm}}l@{\hspace{2.5mm}}c@{\hspace{2.5mm}}c@{\hspace{2.5mm}}c@{\hspace{2.5mm}}c@{\hspace{2.5mm}}c@{\hspace{2mm}}c@{\hspace{2.5mm}}c}
\hline
Binary & Sp. Type & $\log(L/L_\odot)$ & $T_\text{eff}$ [K] & $R_*$ [$R_\odot$] &
$M_*$ [$M_\odot$] & $D$ [$R_\odot$] & $\x L$ [$\text{erg}\,\text{s}^{-1}$]&
$\dot M$ [$M_\odot\,\text{yr}^{-1}$] & \x t & Reference\\
\hline
\object{X Per\tablefootmark{f}} & B0Ve & 4.69 & 32000\tablefootmark{a} & 7.2\tablefootmark{a} & 15.5 & 420 & $ 1.2\times10^{35 }$ &$ 1.5\times10^{-8  }$ & 0.1 & 18, 66 \\
\object{IGR J08408-4503\tablefootmark{f}\tablefootmark{h}} & O8.5I & 5.83 & 34000 & 23.8 & 30 & 59 & $ 3.3\times10^{32 }$ &$ 1.1\times10^{-6  }$ & 2.3 & 50, 51, 52 \\
\object{PSR B1259-63\tablefootmark{f}\tablefootmark{c}} & O9.5Ve & 4.9 & 34000\tablefootmark{d} & 8.1\tablefootmark{d} & 10 & 1000 & $ 3.5\times10^{34 }$ &$ 3.2\times10^{-8  }$ & 0.2 & 21, 22 \\
\object{IGR J16207-5129\tablefootmark{h}} & B1Ia & 5.4 & 29000\tablefootmark{a} & 20 & 20 & 40 & $ 2\times10^{34 }$ &$ 2.2\times10^{-7  }$ & 0.4 & 35, 36 \\
\object{AX J16319-4752} & O8I & 5.63 & 33000 & 20 & 35\tablefootmark{a} & 32 & $ 7.8\times10^{34 }$ &$ 5.1\times10^{-7  }$ & 0.5 & 57, 58 \\
\object{IGR J16328-4726\tablefootmark{h}} & O8Iaf & 5.68 & 33000\tablefootmark{a} & 21.3\tablefootmark{a} & 34.3\tablefootmark{a} & 61 & $ 1.4\times10^{35 }$ &$ 6.2\times10^{-7  }$ & 1.1 & 53, 54 \\
\object{IGR J16418-4532\tablefootmark{h}} & O8.5I & 5.69 & 32800 & 21.7 & 30 & 31.6 & $ 2\times10^{36 }$ &$ 6.4\times10^{-7  }$ & 0.6 & 9, 38 \\
\object{IGR J16465-4507\tablefootmark{h}} & O9.5Ia & 5.69 & 30000\tablefootmark{a} & 26 & 28 & 124 & $ 6.8\times10^{36 }$ &$ 6.4\times10^{-7  }$ & 1.5 & 31, 32 \\
\object{IGR J16479-4514\tablefootmark{h}} & O8.5Iab & 5.52 & 31000\tablefootmark{a} & 20 & 35 & 31.6 & $ 1.09\times10^{34 }$ &$ 3.4\times10^{-7  }$ & 0.3 & 11 \\
\object{IGR J16493-4348} & B0.5Ia & 5.76 & 28000\tablefootmark{a} & 32.2 & 47 & 55 & $ 1.3\times10^{34 }$ &$ 8.2\times10^{-7  }$ & 0.6 & 64, 65 \\
\object{4U 1700-377\tablefootmark{f}} & O6.5Iaf & 5.81 & 35000 & 22 & 46 & 35 & $ 2.2\times10^{36 }$ &$ 1.0\times10^{-6  }$ & 0.9 & 7, 8, 63 \\
\object{IGR J17252-3616} & B0Ia & 5.79 & 30000 & 29 & 15 & 51 & $ 1.6\times10^{37 }$ &$ 9.1\times10^{-7  }$ & 6.9 & 12, 13, 40 \\
\object{IGR J17354-3255\tablefootmark{h}} & O9.5Iab & 5.56 & 30000\tablefootmark{a} & 22.4\tablefootmark{a} & 28.8\tablefootmark{a} & 54 & $ 1.8\times10^{36 }$ &$ 3.9\times10^{-7  }$ & 0.7 & 48, 49 \\
\object{XTE X1739-302\tablefootmark{h}} & O8.5Iab & 5.65 & 32000\tablefootmark{a} & 21.7\tablefootmark{a} & 32.2\tablefootmark{a} & 173 & $ 4.8\times10^{33 }$ &$ 5.4\times10^{-7  }$ & 1.4 & 46, 47 \\
\object{IGR J17544-2619\tablefootmark{h}} & O9Ib & 5.4 & 29000 & 20.0 & 25.9 & 38.0 & $ 1.7\times10^{35 }$ &$ 2.2\times10^{-7  }$ & 0.3 & 33, 34, 56 \\
\object{SAX J1818.6-1703\tablefootmark{h}} & B0.5Iab & 5.7 & 28000\tablefootmark{a} & 30 & 25 & 120 & $ 6\times10^{35 }$ &$ 6.5\times10^{-7  }$ & 1.5 & 61, 62 \\
\object{LS 5039\tablefootmark{f}} & O6.5V & 5.19 & 37500 & 9.3 & 22.9 & 34.5 & $ 6\times10^{34 }$ &$ 9.6\times10^{-8  }$ & 0.3 & 26, 39, 44, 45, 59 \\
\object{IGR J18450-0435\tablefootmark{h}} & O9Ia & 5.58 & 30000 & 23 & 30 & 72 & $ 8\times10^{36 }$ &$ 4.3\times10^{-7  }$ & 0.8 & 16, 17, 43 \\
\object{4U 1907+09} & O8.5Iab & 5.68 & 29760 & 26.2 & 26.0 & 54 & $ 2\times10^{36 }$ &$ 6.2\times10^{-7  }$ & 1.0 & 25 \\
\object{IGR J19140+0951} & B0.5I & 5.47 & 28000\tablefootmark{a} & 23.2\tablefootmark{a} & 25.4\tablefootmark{a} & 70 & $ 3\times10^{35 }$ &$ 2.8\times10^{-7  }$ & 0.6 & 23 \\
\object{Cyg X-1\tablefootmark{f}} & O9.7Iab & 5.57 & 32000 & 19.9\tablefootmark{e} & 24.0 & 42.4 & $ 1.4\times10^{37 }$ &$ 4.1\times10^{-7  }$ & 0.8 & 27, 28, 41 \\
\object{4U 2206+54} & O9.5Ve & 4.7 & 32000\tablefootmark{a} & 7.3 & 16 & 76 & $ 1.8\times10^{35 }$ &$ 1.6\times10^{-8  }$ & 0.1 & 24, 60 \\

\hline
\end{tabular}
\tablefoot{Stellar parameters are taken from the literature, except for the mass-loss
rate, for which we used fits of \citet{cmfkont}, and for the optical depth
parameter, which was calculated from Eq.~\eqref{tx}. \tablefoottext{a}{Derived
using the expressions of \citet{okali}.
\tablefoottext{c}{Possible disk accretion.}\tablefoottext{d}{Polar
values.}\tablefoottext{e}{In the direction of the companion.}}
\tablefoottext{f}{Some alternative designations: X Per (HR 1209),
IGR J08408-4503 (LM Vel, HD 74194),
PSR B1259-63 (CPD-63$\degr$2495),
4U 1700-377 (V884 Sco, HD 153919),
LS 5039 (V479 Sct),
Cyg X-1 (V1357 Cyg, HD 226868),
and  4U 2206+54 (BD+53$^\circ$2790).

}
\tablefoottext{h}{Supergiant fast X-ray transient \citep{lutov,walt,hmxb57}.}
}
\tablebib{(7)~\citet{hmxb7};
(8)~\citet{hmxb8};
(9)~\citet{hmxb9};
(11)~\citet{hmxb11};
(12)~\citet{hmxb12};
(13)~\citet{hmxb13};
(16)~\citet{hmxb16};
(17)~\citet{hmxb17};
(18)~\citet{hmxb18};
(21)~\citet{hmxb21};
(22)~\citet{hmxb22};
(23)~\citet{hmxb23};
(24)~\citet{hmxb24};
(25)~\citet{hmxb25};
(26)~\citet{hmxb26};
(27)~\citet{hmxb27};
(28)~\citet{hadrvitr};
(31)~\citet{hmxb31};
(32)~\citet{hmxb32};
(33)~\citet{hmxb33};
(34)~\citet{hmxb34};
(35)~\citet{hmxb35};
(36)~\citet{hmxb36};
(38)~\citet{hmxb38};
(39)~\citet{hmxb39};
(40)~\citet{hmxb40};
(41)~\citet{hmxb41};
(43)~\citet{hmxb43};
(44)~\citet{hmxb44};
(45)~\citet{hmxb45};
(46)~\citet{hmxb46};
(47)~\citet{hmxb47};
(48)~\citet{hmxb48};
(49)~\citet{hmxb49};
(50)~\citet{hmxb50};
(51)~\citet{hmxb51};
(52)~\citet{hmxb52};
(53)~\citet{hmxb53};
(54)~\citet{hmxb54};
(56)~\citet{hmxb56};
(57)~\citet{hmxb57};
(58)~\citet{hmxb58};
(59)~\citet{hmxb59};
(60)~\citet{hmxb60};
(61)~\citet{hmxb61};
(62)~\citet{hmxb62};
(63)~\citet{hmxb63};
(64)~\citet{hmxb64};
(65)~\citet{hmxb65};
(66)~\citet{hmxb66}.
}
\end{table*}

The parameters of real wind-powered HMXBs should appear outside the region with
wind inhibition (the forbidden region). To test this, we collected parameters of
HMXBs with $\Teff>27\,000\,\text{K}$ from the literature (see
Table~\ref{neutron}). We excluded HMXBs with Roche overflow and Be/X-ray
binaries powered by disk accretion. Positions of collected stars are also
plotted in Figs.~\ref{lxtx} -- \ref{lxtxch10naj}. The X-ray luminosities of
HMXBs were derived from the literature. We used the predicted mass-loss rates
\citep[given also in Table~\ref{neutron}]{cmfkont} and the terminal velocities
derived from the stellar parameters, $v_\infty= 2.6 \,v_\text{esc}=2.6
\hzav{2GM_*(1-\Gamma)/R_*}^{1/2}$ \citep{lsl} to calculate the optical depth
parameter $\x t$ of HMXBs in diagrams of $\lx-\x t$ and in Table~\ref{neutron}.
In Figs.~\ref{lxtxch3} -- \ref{lxtxch10naj} we further corrected the HMXB
mass-loss rates for clumping. We multiplied the rates in Table~\ref{neutron} by
the ratio of mass-loss rates with and without clumping from
Table~\ref{chuchpar}.

Without clumping, many of the stars appear in the region of wind inhibition (see
Fig.~\ref{lxtx}). Contrary to our previous study \citep{dvojvit}, our results
are based on global models, which predict lower mass-loss rates \citep{cmfkont}
and therefore stronger influence of X-rays. As a result of lower mass-loss
rates, some stars moved to the region of wind inhibition.

The inclusion of clumping with $\cc=3$ does not remove stars from the forbidden
region (Fig.~\ref{lxtxch3}). The mass-loss rate of model stars
(Table~\ref{chuchpar}) as well as the predicted mass-loss rate for the stars
from Table~\ref{neutron} increase, causing a shift of the diagram to the
right. The predicted wind terminal velocity becomes lower, increasing $\x t$ for
model stars (see Eq.~\eqref{tx}). Consequently, the location of observed stars
shifts into the region of wind inhibition, and as a result of this, a constant
clumping factor does not improve the agreement between observations and theory.

Stars are mostly removed from the forbidden region only with 
the
%
variable
clumping factor
according to
Eq.~\eqref{najc} (Fig.~\ref{lxtxch10naj}). The mass-loss rates
slightly increase in comparison with the models without clumping (see
Table~\ref{chuchpar}), while the predicted terminal velocity remains large due
to the radial increase of clumping. Moreover, with increasing clumping, the influence
of X-rays becomes lower and as a result of these effects the positions of
observed HMXBs appear outside the region with wind inhibition.

There is direct observational support for the effect of wind inhibition by
X-rays. An O9.5~V star BD+53$^\circ$2790, the optical counterpart to the X-ray
source 4U 2206+54, shows peculiar wind with a very low terminal velocity
\citep{sfalerit}. This star is located close to the border of the wind
inhibition area in Fig.~\ref{lxtxch10naj} and therefore its peculiar wind can be
explained as being the result of X-ray irradiation. Also, X~Per appears outside the area
of the wind inhibition in this diagram, which is not in conflict with the considered
wind accretion in this system \citep{majtra}.

We noticed two stars (not listed in Table~\ref{neutron}) that may be located
in the area of wind inhibition. For 4U 1538-522, the parameters from \citet{byk}
predict $\x t\approx0.2$ for $\lx= 2.9\times10^{36}\,\ergs$. A similar problem
appears for XTE J1855-026 and parameters from \citet{hmxb63}. These systems are
very compact (distance between the neutron star and the primary is only a small
fraction of the radius of the primary) and, possibly, the X-rays are powered by
Roche-lobe overflow and not by wind accretion. An alternative explanation is
that the assumption of a simple geometry of the problem (see Fig.~\ref{geom})
breaks down for compact systems.

\subsection{Wind inhibition in the models with
radial dependent clumping: wind parameters and modification of CAK line force}

Because the models with radial dependence of a clumping factor according to
Eq.~\eqref{najc} (following \citealt{najradchuch}) provide the best match with observations, we focus here on models with this approximation of clumping. The decrease of the wind
terminal velocity due to X-ray irradiation can be roughly described by a formula
that resembles wind $\beta$-law (see Fig.~\ref{ch10najventau}),
\begin{equation}
\label{ch10najventaurov}
v_\infty(\lx,D)=v_{\infty,0}\zav{1-\frac{R_*}{D}}^{\beta_1(\lx/L_{36})^{\beta_2}},
\end{equation}
where $v_{\infty,0}$, $\beta_1$, and $\beta_2$ are fit parameters. The values
of these parameters derived by fitting model terminal velocities as a function
of $\lx$ and $D$ are given in Table~\ref{ch10najventautab}. Here
$L_{36}=10^{36}\,\ergs$.

The dependence of the model
wind mass-loss rate on the X-ray luminosity and on the
binary separation is slightly more complicated. The mass-loss rate first
slightly increases with increasing \lx\ (by the order of 10\%) due to increased
radiative force from newly appearing ionization states. However, for even
stronger \lx\ the mass-loss rate decreases leading to wind inhibition. The
behaviour of the mass-loss rate close to the limit of wind inhibition can be
very roughly approximated as
\begin{equation}
\label{ch10najventaurovdmdt}
\dot M(\lx,D)=\dot M_0\hzav{1-\exp\zav{-\frac{\zav{D-R_*}^2}{s_1(\lx/L_{36})^{s_2}}}},
\end{equation}
where $\dot M_0$, $s_1$, and $s_2$ are fit parameters. The values
of these parameters derived by fitting predicted mass-loss rates as a function
of $\lx$ and $D$ are given in Table~\ref{ch10najventautab}.

Our models were calculated using X-ray irradiation in the radial direction
(assuming zero inclination, Fig.~\ref{geom}) but in reality the inclination of
incident X-rays might be non-zero \citep[e.g.][]{velax1}. As a result, the
terminal velocity and the mass-flux depend on direction. Therefore,
Eq.~\eqref{ch10najventaurov} predicts only the terminal velocity in the
direction of the compact companion and the quantity  $\dot M(\lx,D)/(4\pi)$
(see Eq.~\ref{ch10najventaurovdmdt}) gives the mass-flux per unit solid angle in
this direction.

\begin{table}[t]
\caption{Derived parameters of the terminal velocity fit~\eqref{ch10najventaurov} and mass-loss rate fit
\eqref{ch10najventaurovdmdt} for individual models with X-ray irradiation.}
\label{ch10najventautab}
\centering
\begin{tabular}{ccccccc}
\hline
Model & $v_{\infty,0}$ & $\beta_1$ & $\beta_2$ & $\dot M_0$ & $s_1$ &
$s_2$\\
 & [\kms] & & & $[\text{M}_{\odot}\,\text{yr}^{-1}$] \\
\hline
300-1 & 1720 & 2.31 & 0.337 & $6.8\times10^{-7}$ & 196  & 0.758\\
375-1 & 2340 & 1.16 & 0.299 & $2.4\times10^{-6}$ & 20.4 & 0.910\\
300-5 & 2280 & 8.22 & 0.452 & $2.7\times10^{-8}$ & 200  & 0.754\\
375-5 & 3410 & 3.90 & 0.376 & $2.3\times10^{-7}$ & 51.4 & 0.729\\
\hline
\end{tabular}
\end{table}

\begin{table}
\caption{Line force multipliers for the models with clumping following
Eq.~\eqref{najc}.}
\centering
\label{cakpar}
\begin{tabular}{ccccc}
\hline
Model & $k$ & $\alpha$ & \vrule height2.0ex width0pt $\bar Q$ &
$\xi_\text{kink}$ [$\text{erg}\,\text{cm}\,\text{s}^{-1}$] \\
\hline
300-1 & 0.099 & 0.575 & 220 & 9.2\\
375-1 & 0.090 & 0.620 & 630 & 12.6\\
300-5 & 0.520 & 0.410 & 100 & 22.8\\
375-5 & 0.135 & 0.580 & 470 & 19.2\\
\hline
\end{tabular}
\end{table}

In Table \ref{cakpar} we provide the radiative force multipliers $k$ and
$\alpha$ \citep{cak,abpar} corresponding to models with radially dependent
clumping after Eq.~\eqref{najc}. These force multipliers can be used to
approximate the radiative force in hydrodynamical simulations without solving
the NLTE problem. The force multipliers describe the strength ($k$) and slope
($\alpha$) of the line distribution function \citep{pusle}, but we simply
selected the multipliers that give the best match with wind mass-loss rate and
terminal velocity of models without X-ray irradiation (given in
Table~\ref{chuchpar}). We also provide the $\bar Q$ parameter of \citet{gayley},
which is more substantially physically motivated and is related to the remaining
force multipliers as $\bar
Q=\hzav{(1-\alpha)k\zav{c/v_\text{th}}^\alpha}^{1/(1-\alpha)}$, where the
fiducial hydrogen thermal speed is $v_\text{th}=\sqrt{2 k_\text{B}
T_\text{eff}/m_\text{H}}$).

The effect of X-ray irradiation can be roughly included in hydrodynamical models
by multiplication of the CAK line force (Eq.~(6) of \citealt{cak} or Eq.~(16) of
\citealt{ppk}) by a factor of $\exp(-\xi/\xi_\text{kink})$. Here, $\xi$ can be
calculated from Eq.~\eqref{xic} and the mean values of $\xi_\text {kink} $ for
individual model stars are given in Table~\ref{cakpar}.

\section{Wind-powered X-ray luminosity}
\label{lxsec}

\begin{figure}[t]
\centering
\resizebox{\hsize}{!}{\includegraphics{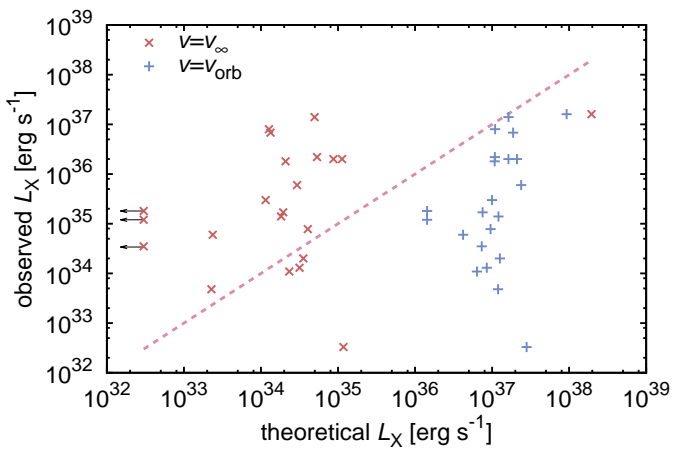}}
\caption{Observed X-ray luminosity in comparison with predicted X-ray
luminosity~\eqref{lxlxrov} for the stars from Table~\ref{neutron} for two
limiting values of relative velocity: $v_\infty$ (red crosses)
and $v_\mathrm{orb}$ (blue plus signs).
Line denotes one-to-one relation.}
\label{lxlx}
\end{figure}

The X-ray luminosity of wind-powered HMXBs stems from the release of the
gravitational potential energy during wind accretion. Within the classical
Bondi-Hoyle-Lyttleton picture \citep{holy,boho} the accretion luminosity is
\begin{equation}
\label{lxlxrov}
\lx=\frac{G^3\x M^3}{\x R D^2 v^4}\dot M,
\end{equation}
where $\x M$ and $\x R$ are the mass and radius of an accreting object, and $v$ is
its relative velocity, which can be estimated using the  orbital velocity of the
compact component $v_\mathrm{orb}$ and the wind velocity at the distance
$D$ of the compact component, $v_\mathrm{wind} = v(D)$ as
\begin{equation}
\label{pytha}
v^2=v_\mathrm{wind}^2+v_\mathrm{orb}^2.
\end{equation}
There are two limiting cases: either $v=v_\infty$ when the wind velocity is not
affected by the X-ray source and the wind is accreted at large distances from
the non-degenerate star, or $v=v_\mathrm{orb}$ in an opposite case. Either way, for
a given set of system parameters Eq.~\eqref{lxlxrov} gives the maximum X-ray
luminosity assuming full conversion of gravitational potential energy to X-rays.

\citet{sandvelax} give a more precise estimate of X-ray luminosity, where
instead of the $1/v^4$ dependence in Eq.~\eqref{lxlxrov}, they introduce a dependence of 
$1/(v^3v_\mathrm{wind})$. However, with this formula the X-ray
luminosity formally diverges, $\lx\rightarrow\infty$ for
$v_\mathrm{wind}\rightarrow0$. This is avoided in reality, because for a very
low $v_\mathrm{wind}$ the wind becomes fully inhibited, does not reach the
gravitational well of the compact star and is trapped in the gravitational well
of a donor star. Therefore, the wind velocity should be at least roughly
$v_\mathrm{orb}$ to reach the compact star. To account for this effect at least
approximately, we keep the $1/v^4$ dependence in Eq.~\eqref{lxlxrov} which does not
diverge for $v_\mathrm{wind}\rightarrow0$.

The relation between observed and predicted X-ray luminosity according to
Eq.~\eqref{lxlxrov} for stars from Table~\ref{neutron} is plotted in
Fig.~\ref{lxlx}. Without X-ray wind inhibition (for $v=v_\infty$), most of the
stars are located above the theoretical relation. Although this might be partly
due to the fact that the compact object is located close to the massive star
where the wind has not yet reached the terminal velocity \citep{sandvelax}, for
many objects this would likely mean that their X-ray emission could not be
wind-powered. The agreement cannot be significantly improved by increasing the
mass-loss rate, because from Eq.~\eqref{lxlxrov} the X-ray luminosity scales
with mass-loss rate only linearly and an increase of the mass-loss rate by two
orders of magnitude (with respect to current models) is not justified.

On the other hand, all stars are located below the maximum theoretical
expectation for the case of extreme X-ray wind inhibition (for $v=v_\mathrm{orb}
= \sqrt{GM_*/D}$, see Fig.~\ref{lxlx} and also \citealt{hohoho}). This is caused
by a very strong dependence, $\lx\sim v^{-4}$. This shows that current wind
mass-loss rate predictions are able to explain the observed X-ray luminosities
in HMXBs. Moreover, this is strong observational support for the existence of
the wind velocity decrease due to X-ray radiation. Most stars do not reach the
maximum X-ray luminosity. There may be several reasons for this: for many stars
$v>v_\text{orb}$,  for a given star there is a maximum X-ray luminosity
due to wind inhibition, and finally the accretion may be suppressed by the
compact star itself \citep{nerozumim}.

\begin{figure}[t]
\centering
\resizebox{\hsize}{!}{\includegraphics{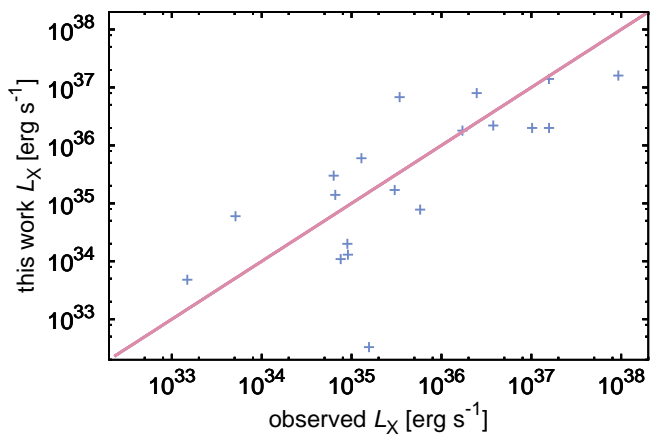}}
\caption{X-ray luminosities
of stars from Table~\ref{neutron}
estimated using 
Eqs.~\eqref{lxlxrov} and \eqref{pytha} inserting the terminal
velocity
(as $v_\mathrm{wind}$)
derived from Eq.~\eqref{ch10najventaurov}
using
observed X-ray
luminosities. Plotted as a function of observed X-ray luminosity.
Line denotes one-to-one relation.}
\label{lxlxv}
\end{figure}

The decrease of the wind terminal velocity due to X-ray irradiation predicted by
Eq.~\eqref{ch10najventaurov} can be used to check the consistency of estimated
X-ray luminosities with observations. Inserting $v_\infty(\lx,D)$ from
Eq.~\eqref{ch10najventaurov} calculated for observed X-ray luminosity and scaled
by the ratio of the terminal velocity of a given star (after \citealt{lsl}) and
model terminal velocity from Table~\ref{chuchpar} (for $C_1=10$) into
Eq.~\eqref{pytha}, we obtain from Eq.~\eqref{lxlxrov} an estimate of X-ray
luminosity. These estimates nicely match the observed X-ray luminosities (see
Fig.~\ref{lxlxv}).

\begin{figure}[t]
\centering
\resizebox{\hsize}{!}{\includegraphics{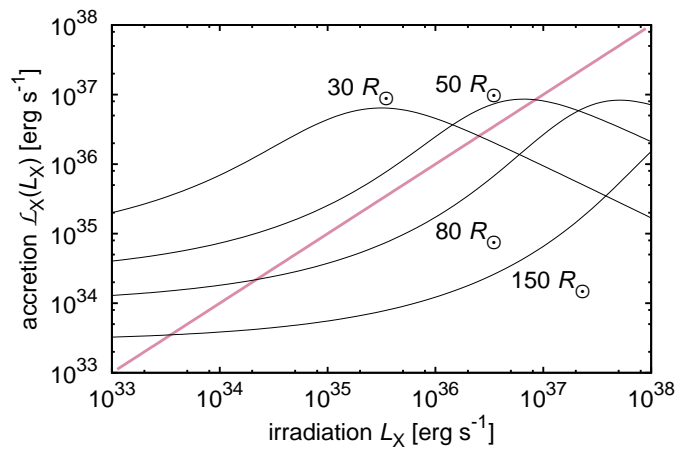}}
\caption{Accretion X-ray luminosity according to Eq.~\eqref{lxlxrovimp} as a
function of irradiation X-ray luminosity for the model 300-1 (black lines).
Labels denote binary separation $D$. The intersections of straight red line,
which denotes one-to-one relation, with black curves correspond to the solution
of Eq.~\eqref{lxlxrovimp}.}
\label{lxlxlx}
\end{figure}

Inserting the terminal velocity fit~\eqref{ch10najventaurov} and mass-loss
rate fit~\eqref{ch10najventaurovdmdt} into Eq.~\eqref{lxlxrov} for a given
$D$, we derive an equation for the accretion X-ray luminosity as a function of
X-ray irradiation luminosity
\begin{equation}
\label{lxlxrovimp}
\lx^\mathrm{acc}=\x {{\cal L}}(\lx^\mathrm{irr})=\frac{G^3\x M^3}{\x R D^2}
\frac{\dot M(\lx,D)}{\zav{v^2_\infty(\lx^\mathrm{irr},D)+v_\mathrm{orb}^2}^2}.
\end{equation}
The irradiating X-rays modify the wind terminal velocity and mass-loss rate, and
therefore the accretion luminosity $\lx^\mathrm{acc}$ is a function of the
irradiation luminosity $\lx^\mathrm{irr}$ (Fig.\,\ref{lxlxlx}). For small
irradiation luminosities the X-rays do not significantly influence either the terminal
velocity or the mass-loss rate. Consequently, the accretion luminosity is
almost constant. For higher irradiation luminosities the terminal velocity
decreases, which results in an increase of the accretion luminosity. The curves
reach a maximum accretion luminosity and start to decrease due to the decrease of
the wind mass-loss rate with X-ray luminosity for the case of strong X-ray
irradiation. This is nicely illustrated for orbital separations $30\,R_*$ and
$50\,R_*$ in Fig.\,\ref{lxlxlx}. With decreasing orbital separation the
influence of X-rays becomes stronger, which results in the shift of the maximum
to lower irradiation luminosities.

In a stationary state the irradiation and accretion luminosities are equal,
$\lx^\mathrm{irr} = \lx^\mathrm{acc}$, and are given as a solution of
Eq.~\eqref{lxlxrovimp}. This is an implicit equation, therefore the solution has
to be found numerically. The solution of Eq.~\eqref{lxlxrovimp} can be obtained
graphically from Fig.~\ref{lxlxlx} using iterations $L_\text{X,n+1}= \x {{\cal
L}} (L_\text{X,n})$ for $n\rightarrow\infty$. According to the Banach
fixed-point theorem \citep{banasarna}, such iterations converge if $|\de \x
{{\cal L}}/\de \lx |<1$. This holds for all curves in Fig.~\ref{lxlxlx}
and consequently we can expect that the X-ray emission should be stable even in the
presence of X-ray feedback. Therefore, proper treatment of X-ray feedback may
suppress large fluctuations of mass-accretion seen in numerical simulations
\citep{blondyn}. Moreover, as a result of the shape of the curves, there are two
types of solution with $\lx=10^{33}-10^{34}\,\ergs$ and
$\lx=10^{36}-10^{37}\,\ergs$ \citep{hohoho}.

This dichotomy may provide a natural explanation for the two classes of X-ray
binaries, that is, the supergiant fast X-ray transients (SFXTs) and classical
supergiant X-ray binaries (sgXBs). The two classes of sgXBs
have very similar binary parameters, but differ in strength and variability of
their X-ray emission. The X-ray emission of sgXBs is several orders of magnitude
stronger on average, while SFXTs show fast and strong variability. This can be
explained either by the difference in the wind accretion \citep{nerozumim} or by
the difference in the wind itself \citep{hmxb56,praboz}. The latter explanation
is supported by the systematic difference between the absorbing column densities
of SFXTs and sgXBs \citep{praboz}. Based on our models, the differences
between sgXBs and SFXTs can be explained by the strength of the X-ray feedback. In
a state with low X-ray luminosity, the wind is basically unaffected by X-rays,
has a relatively high speed, and from the continuity equation $\dot M=4\pi
r^2\rho v$ has a very low density $\rho\sim 1/v$. As a result of the high speed, the
accreted amount of mass is low (see Eq.~\eqref{lxlxrov}) and as a result of low
density the absorbing column is also low,
which are characteristic properties of SFXTs.
On the other
hand, with large X-ray luminosity the wind velocity decreases, increasing the
mass-accretion rate and the absorbing column density.
The latter
corresponds to sgXBs.

The parameters of SFXTs in Table~\ref{neutron} (denoted with index $h$ there)
support this picture. Either the optical depth parameter of SFXTs is typically
large, $\x t\gtrsim1$, or the binaries are located in the bottom part of $\lx-\x
t$ diagrams.

From Fig.~\ref{lxlxlx} it follows that just a small change of binary separation
on the order of ten percent may cause transition from the SFXT regime to the sgXB
regime. Another parameter that may determine the type of the object is the
stellar luminosity, by setting of the wind mass-loss rate or the surface gravity,
which determines the terminal velocity. In our explanation we neglected the role
of the magnetic field and spin of the neutron star, which may still be important in
affecting the mode of accretion \citep{nerozumim,bozof}.

The existence of X-ray luminosity maxima in Fig.~\ref{lxlxlx} is connected with
the self-regulated state found in sgXBs \citep{dvojvit}. For irradiation X-ray
luminosities lower than the maximum one, an increase of the X-ray luminosity
leads to stronger feedback, lower wind velocities, and stronger X-ray
emission. On the other hand, if the irradiation X-ray luminosities overshoot the maximum accretion X-ray luminosity, this leads to a decrease of the
wind mass-loss rate and a decrease of the accretion X-ray luminosity. 

Sensitivity of wind feedback on orbital and wind parameters may also contribute
to the variability of X-ray luminosity on flow timescale
$\tau_\text{flow}\approx D/v_\infty$, which is of the order of hours to days.
The transitions between states with high and low X-ray luminosity may be
triggered by the wind perturbations that include inhomogeneities on small or
large scale \citep{osfek,duo}.

\section{Conclusions}

We studied wind inhibition in wind-powered HMXBs. We included the X-ray
irradiation into our wind models and studied its effect on the radiative force
for various X-ray luminosities and binary separations. As a result of strong
X-ray irradiation, the radiative force that drives the wind decreases. This
leads to the appearance of a kink in the wind velocity, which reduces the wind
terminal velocity. For a very strong X-ray irradiation, the position of the kink
approaches the star and the wind may fall back onto the star or the mass-loss
may even reduce, leading to wind inhibition.

This wind inhibition may be conveniently studied in diagrams that combine the
X-ray luminosity and the optical depth parameter. There was a good agreement
between the position of observed stars in the \lx~versus~$\x t$ diagrams and the
theoretical expectations in our previous models \citep{dvojvit}. However, more
advanced global wind models predict significantly lower wind mass-loss rates,
which results in the appearance of stars in the area corresponding to the wind
inhibition.

The above agreement can be improved by introducing optically thin clumping
(microclumping). Small-scale wind inhomogeneities (clumping) are expected to
appear in the winds based on both theory and empirical evidence. Clumping
improves the agreement between the expected and observed positions of the stars in
the \lx~versus~$\x t$ diagrams, because it weakens the effect of X-ray irradiation
(as it favours recombination) and leads to an increase of the wind mass-loss
rate. We tested different types of radial variations of the clumping factor and
the best match between the models and observed properties of HMXBs was derived
with the radially variable clumping of \citet{najradchuch}.

Based on our models, we described the influence of X-ray irradiation on the
terminal velocity and on the mass-flux in the direction of a compact star in a
parametric way. This enabled us to estimate the X-ray luminosities of individual
HMXBs from their binary parameters assuming Bondi accretion. Without introducing
the wind inhibition, the expected X-ray luminosities are typically lower than observed values by two to three
orders of magnitude. However, introducing the
reduction of wind velocities due to X-rays, the compact companion accretes
a much larger amount of the wind and produces larger X-ray luminosities in
agreement with observations.

As a result of the ionizing feedback of X-ray radiation, the equation for the
X-ray luminosity within the Bondy-Hoyle-Lyttleton accretion picture becomes
implicit. The derived roots however yield a stable solution of this equation.
This also implies that the resulting X-ray luminosity can be expected to be
stable. Moreover, there are two types of solution. One solution corresponds to
weak X-ray irradiation with $\lx=10^{33}-10^{34}\,\ergs$ and the second solution
corresponds to $\lx=10^{36}-10^{37}\,\ergs$. These two types of solution may
provide a natural explanation for the two classes of X-ray binaries, that is, the
SFXTs and sgXBs.

We also provide CAK line force multipliers for studied stars and introduce
additional parameters that account for the influence of X-ray irradiation. These
parameters can be used in hydrodynamical simulations.

\begin{acknowledgements}
We thank Dr.~J.~Puls for comments on the paper and for the discussion on the
inclusion of clumping in the wind models.
This research was supported by grant GA\,\v{C}R 18-05665S.
Computational resources were provided by the CESNET LM2015042 and the CERIT
Scientific Cloud LM2015085, provided under the programme "Projects of Large
Research, Development, and Innovations Infrastructures".
The Astronomical Institute Ond\v{r}ejov is
supported by a project \mbox{RVO:67985815} of the Academy of Sciences of the
Czech Republic.
\end{acknowledgements}

\newcommand\elba{1995, IAU Symp. No. 163, Wolf-Rayet stars: binaries; colliding
winds; evolution, eds. K. A. van der Hucht \&  P. M. Williams (Dordrecht: Kluwer
Academic Publishers)}

\appendix

\section{Implementation of clumping into the wind code}
\label{meluzina}

\begin{figure*}[t]
\centering
\resizebox{0.49\hsize}{!}{\includegraphics{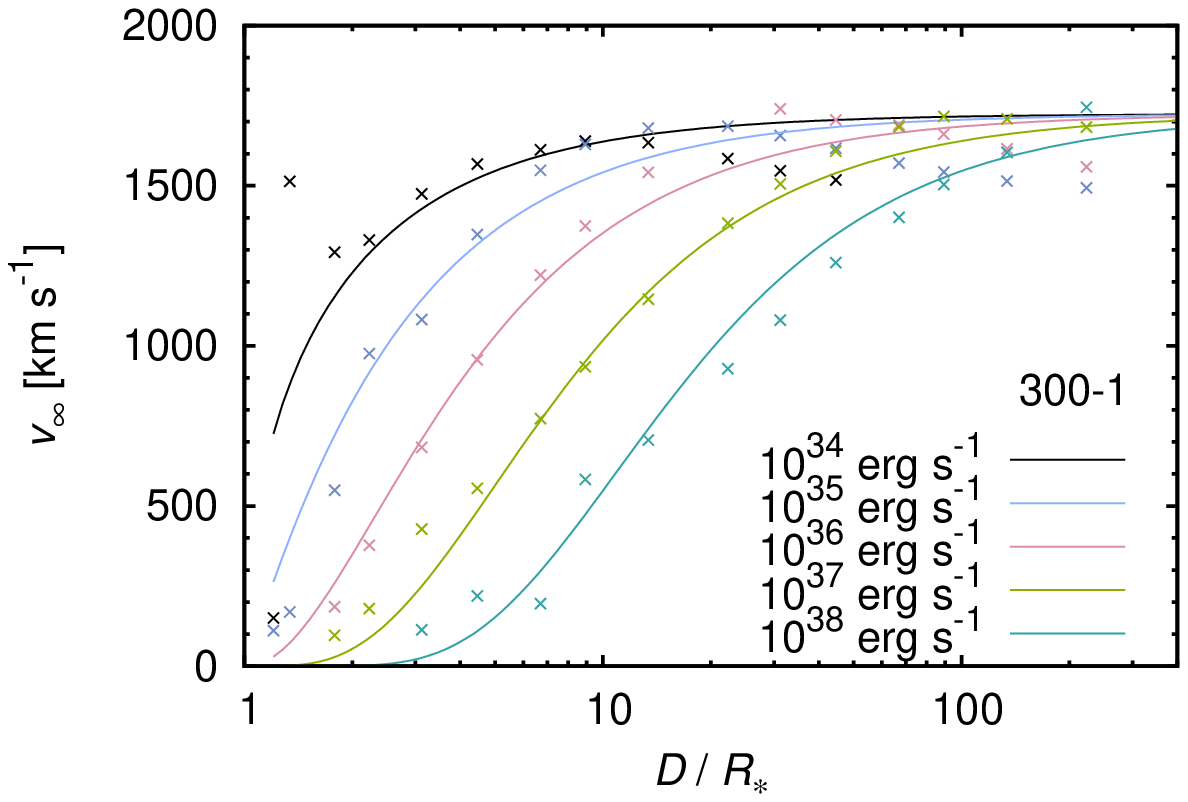}}
\resizebox{0.49\hsize}{!}{\includegraphics{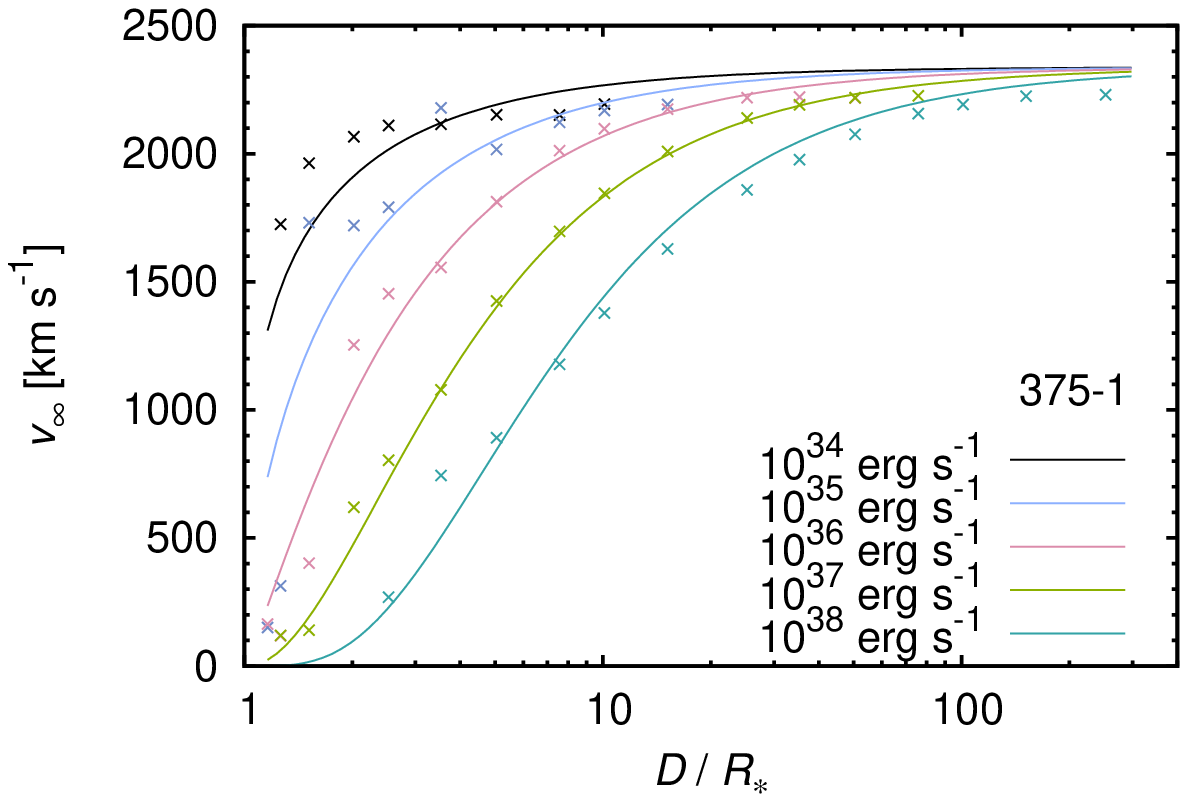}}
\resizebox{0.49\hsize}{!}{\includegraphics{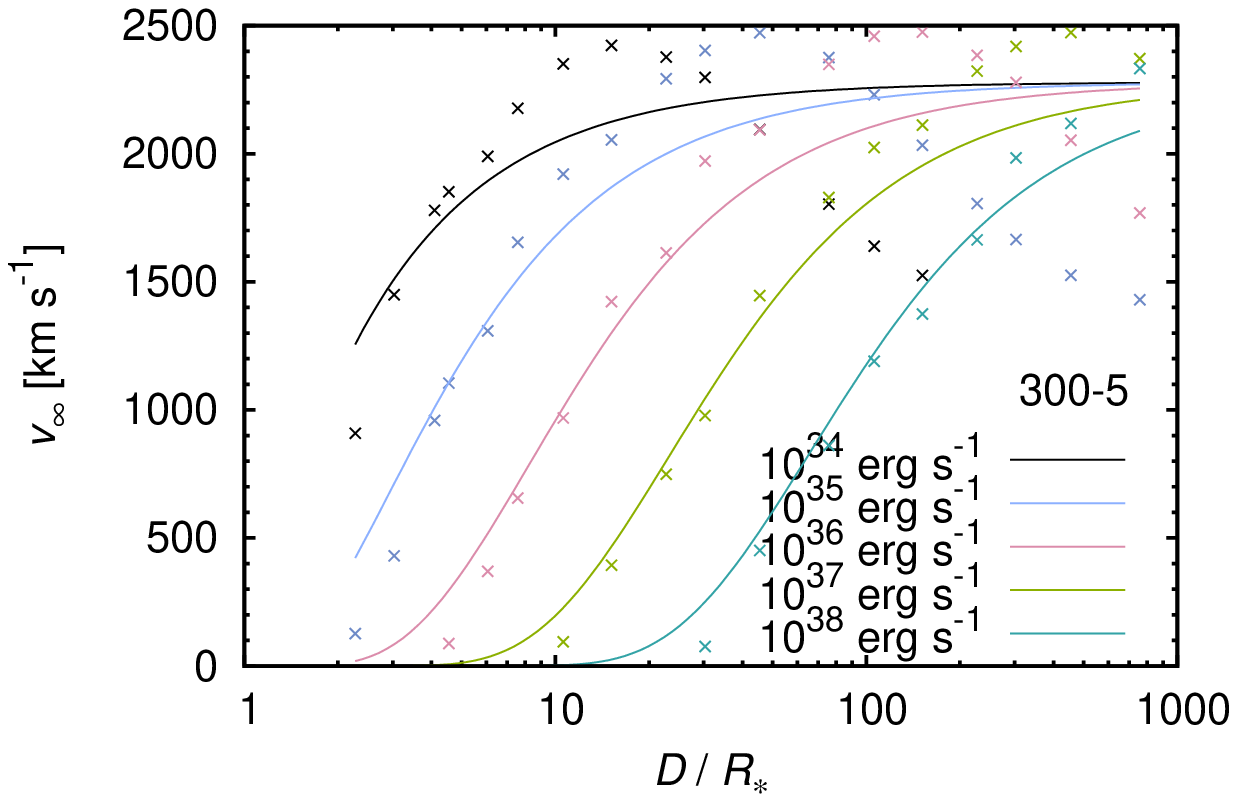}}
\resizebox{0.49\hsize}{!}{\includegraphics{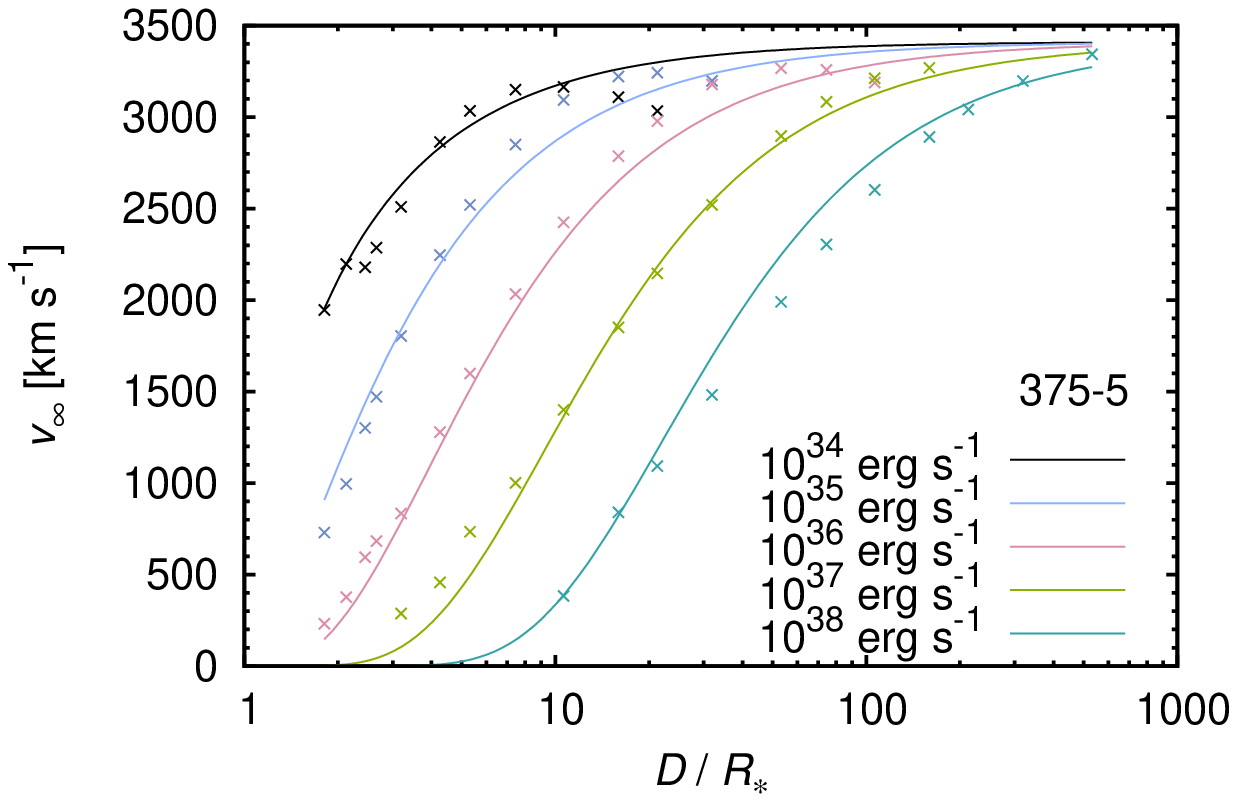}}
\caption{Dependence of the terminal velocity on the binary separation for models
from Table~\ref{ohvezpar} with clumping after Eq.~\eqref{najc} for different
X-ray luminosities (crosses). Overplotted is the corresponding fit
Eq.~\eqref{ch10najventaurov} for different values of \lx.}
\label{ch10najventau}
\end{figure*}

The mean opacity $\prm\chi$ in the case of a clumpy medium can be derived
from the contributions of clumps ($+$) and an interclump medium ($-$),
\begin{equation}
\label{chic}
\prm\chi=\int^{\,(+)}\chi\cpl\,\de V+\int^{\,(-)}\chi\cmi\,\de V=
f\chi\cpl=f\kappa\cpl\rho\cpl=\kappa\cpl\prm\rho,
\end{equation}
where $\kappa\cpl$ is the opacity per unit of mass in the clumps,  $(+)$ and $(-)$
denote the integration in the clumps only and in the interclump media
only, respectively, and where we integrate over a suitably chosen volume $V$ containing a large
number of clumps.
We assumed a void interclump medium, and consequently $\chi\cmi=0$.
Using the mean opacity it is possible to calculate the optical depth,
\begin{multline}
\label{tauc}
\tau=\int\kappa\rho\,\de r = \int^{\,(+)}\kappa\cpl\rho\cpl\,\de
r+\int^{\,(-)}\kappa\cmi\rho\cmi\,\de r=\\*
\frac{1}{f}\int^{\,(+)}\kappa\cpl\prm\rho\,\de r=
\int\kappa\cpl\prm\rho\,\de r = f\int\kappa\cpl\rho\cpl\,\de r=
\int\prm\chi\,\de r,
\end{multline}
where we used Eqs.~\eqref{rhor}, \eqref{rhos}, and the definition of $f$.
The volume-averaged emissivity is
\begin{equation}
\label{etac}
\prm\eta=\int^{\,(+)}\eta\cpl\,\de V+\int^{\,(-)}\eta\cmi\,\de V=
f\eta\cpl,
\end{equation}
where we again used the definition of $f$.

In our models,
we solve for the relative occupation numbers $N_i=n_i/n_\text{elem}$,
where $n_i$ is the occupation number of a level $i$ and $n_\text{elem}$ is the
number density of a given element \citep{nltei}. Consequently, to include the
clumping we modified our code in a following way:
\begin{itemize}
\item The electron number density that features in the statistical
equilibrium equations (and consequently also in the calculation of the
Saha-Boltzmann factors) is taken as $\cc\langle n_\text{e}\rangle$.
\item The electron number density that features in the free-free emission term is
also taken as $\cc\langle n_\text{e}\rangle$.
\item The electron number density that features in the equations of thermal
balance of electrons is taken as $\cc\langle n_\text{e}\rangle$.
\end{itemize}
These modifications enable us to take into account the optically thin clumps. For
example, using the relative number densities inside the clumps we calculate the
mean opacity (see Eq.~\eqref{tauc}) as $\prm\chi=\kappa\cpl\prm\rho$. The
modified Saha-Boltzmann factors enable us to calculate correct free-bound
emissivity in a clumpy environment (see Eq.~\eqref{etac}). The mentioned
modification of the electron density is necessary to obtain correct emissivity
due to the free-free transitions. Finally, with using the relative number
densities $N_i$ and the mean wind density it is possible to obtain the correct
opacity for the solution of the radiative transfer equations in lines (which is
used in the statistical equilibrium equations and for the calculation of the
radiative force).

We applied a test devised by J.~Puls (priv.~commun.) to check for the
validity of including clumping. Our code successfully passed the test.
Moreover, the ionization parameter Eq.~\eqref{xic} also fulfills the test.
\end{document}